\documentclass[10pt,twocolumn,letterpaper]{article}

\usepackage[pagenumbers]{cvpr} %

\input{preamble}

\usepackage{amsmath,amssymb,amsfonts,mathtools,amsthm}
\usepackage{algorithm}
\usepackage{algpseudocode}
\usepackage{graphicx}
\usepackage{textcomp}
\usepackage{multicol}
\usepackage{multirow}
\usepackage{tabularx}
\usepackage{makecell}
\usepackage{xcolor}
\usepackage{xspace}
\usepackage{threeparttable}
\usepackage{fancyhdr}
\usepackage{paralist}
\usepackage[normalem]{ulem}  %
\usepackage{microtype}
\usepackage{subcaption}
\usepackage{booktabs}         %
\usepackage{xurl}
\usepackage{nicefrac}
\newlist{myitemize}{itemize}{1}
\setlist[myitemize]{label=\textbullet, leftmargin=*, itemsep=0pt, topsep=0pt, parsep=0pt, partopsep=0pt}
\usepackage{longtable,booktabs}
\usepackage{etoc}
\usepackage{placeins}
\usepackage[symbol]{footmisc}

\definecolor{cvprblue}{rgb}{0.21,0.49,0.74}
\usepackage[pagebackref,breaklinks,colorlinks,allcolors=cvprblue]{hyperref}

\def\paperID{11860} %
\def\confName{CVPR}
\def\confYear{2026}

\title{\tool: Radioactive Multi-Bit Semantic-Latent Watermarking \\ for Diffusion Models}

\author{
Kexin Li\textsuperscript{1} \quad
Guozhen Ding\textsuperscript{1} \quad
Ilya Grishchenko\textsuperscript{1} \quad
David Lie\textsuperscript{1}\\[0.5em]
\textsuperscript{1}University of Toronto
}

\begin{document}

\twocolumn[{%
\renewcommand\twocolumn[1][]{#1}%
\maketitle

\begin{center}
\captionsetup{type=figure}

\begin{minipage}[b]{0.3\textwidth}
    \centering
    \hspace{1em} \textbf{Original} \hspace{2em} \textbf{Watermarked}\\[0pt]
    \includegraphics[width=0.49\linewidth]{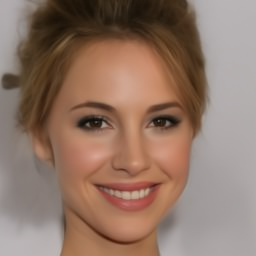}
    \includegraphics[width=0.49\linewidth]{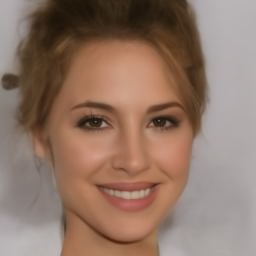}
\end{minipage}
\hspace{1em}
\begin{minipage}[b]{0.3\textwidth}
    \centering
    \hspace{1em} \textbf{Original} \hspace{2em} \textbf{Watermarked}\\[0pt]
    \includegraphics[width=0.49\linewidth]{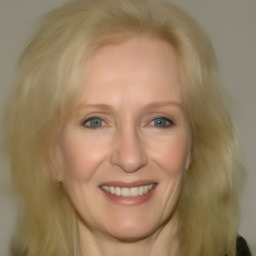}
    \includegraphics[width=0.49\linewidth]{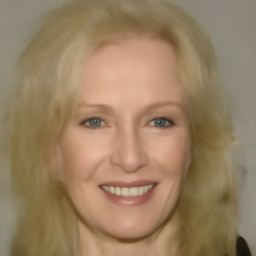}
\end{minipage}
\hspace{1em}
\begin{minipage}[b]{0.3\textwidth}
    \centering
    \hspace{1em} \textbf{Original} \hspace{2em} \textbf{Watermarked}\\[0pt]
    \includegraphics[width=0.49\linewidth]{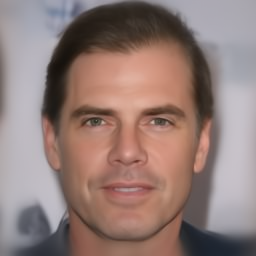}
    \includegraphics[width=0.49\linewidth]{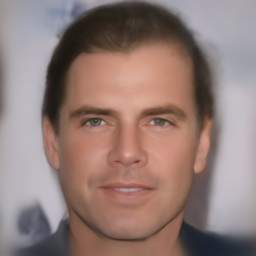}
\end{minipage}

\captionof{figure}{Visual comparison between original and watermarked images.
Each pair shows the original (left) and watermarked (right) images.}
\label{fig:orig_wm_pairs}
\end{center}
}]

\begin{abstract}

Modern generative diffusion models rely on vast training datasets, often including images with uncertain ownership or usage rights. Radioactive watermarks---marks that transfer to a model's outputs---can help detect when such unauthorized data has been used for training.
Moreover, aside from being radioactive, an effective watermark for protecting images from unauthorized training also needs to meet other existing requirements, such as imperceptibility, robustness, and multi-bit capacity.
To overcome these challenges, we propose \tool, a novel multi-bit watermarking scheme, which encodes ownership information as secret bits in the semantic-latent space (h-space) for image diffusion models. 
By leveraging the interpretability and semantic significance of h-space, ensuring that watermark signals correspond to meaningful semantic attributes, the watermarks embedded by \tool exhibit radioactivity, robustness to distortions, and minimal impact on perceptual quality. 
Experimental results demonstrate that \tool achieves 98.57\% watermark detection accuracy, 95.07\% bit-level recovery accuracy, 100\% recall rate, 
and 1.0 AUC on images produced by the downstream adversarial model finetuned with LoRA on watermarked
data across various types of distortions.

\end{abstract}

\setlength{\footnotemargin}{2.5pt}
\footnotetext{$^{*}$Under review.}

\section{Introduction} \label{sec:intro}

Diffusion models (DMs) produce high-quality photo-realistic images
~\cite{rombach2022ldm,ramesh2022hierarchicaltextconditionalimagegeneration,midjourney, saharia2021imagesuperresolutioniterativerefinement,saharia2022photorealistictexttoimagediffusionmodels}.
Such capabilities are not achievable without training on massive datasets of images. 
Unfortunately, this hunger for training data means that model builders do not always obtain the necessary consent and authorization to use data.
This practice has raised legal and ethical concerns, particularly when the images generated by the trained models closely resemble those of artists and studios, undermining originality and creativity. 
For instance, Studio Ghibli, renowned for its distinctive animation style, has become a frequent target of stylistic imitation through publicly accessible generative platforms (\eg, OpenAI)~\cite{bylo2025ghibli,fotor2025ghibli,chatgpt2025}. 
Moreover, methods such as LoRA finetuning~\cite{hu2021loralowrankadaptationlarge} and personalization~\cite{ruiz2023dreamboothfinetuningtexttoimage,textualinversion} enable a DM to learn an artist's style (\ie, style transfer) with only a few images.
Disputes over whether unauthorized training has taken place and whether the outputs constitute a violation of IP rights have resulted in high-profile lawsuits, such as \textit{Andersen v. Stability AI} and \textit{Getty Images v. Stability AI} 
~\cite{apnews2024ghibli,san2025ghibli,wjlta2025ghibli,ippress2025ghibli,jipel2025andersen,artlaw2024andersen,verge2023huckabee,legaldive2024millette,getty_v_stability_complaint,apnews_getty_stability,reuters_getty_stability}.
A key consideration in such disputes is proving that unauthorized images were indeed used in the training of the model.

Image \emph{watermarks} are important tools in the arsenal that content producers have to protect their IP against unauthorized training on generative models~\cite{luo2025digital,min2024watermarkdiffusion,zhang2025aigcwatermark}. Watermarking involves manipulating an image to embed a specially crafted signal (\ie, a watermark). A corresponding detection mechanism can then distinguish between images that contain an embedded watermark and those that do not. Thus, detecting that the output of DM contains a watermark provides strong evidence that the DM trained on watermarked images.
For this purpose, a watermark must satisfy four key properties:
(1) \textit{robustness}: the watermark should not be removable without severely damaging the quality of the watermarked image;
(2) \textit{fidelity}: 
the watermark needs to preserve the perceptual quality of the protected image;
(3) \textit{multi-bit capacity}: 
the watermark should not only be detectable (zero-bit) but also encode ownership information (multi-bit capacity), \eg, to support traitor tracing~\cite{chor1994tracing};
and (4) \textit{radioactivity}~\cite{arewmradioactive}: the watermark should be present in the output of a DM trained on watermarked images.
The last property, in particular, is not sufficiently addressed in the literature. 

Existing watermarking schemes can be broadly categorized into three types: pixel-space, modifying the pixels of the image directly; latent-space, modifying the latent space of a trained model; and semantic watermarks, altering generated images in a human-interpretable way. Pixel-space watermarks have been shown to be non-robust and can be easily removed without damage to the image~\cite{zhao2024invisibleimagewatermarksprovably}. Similarly, latent-space watermarks can be non-robust and also tend to be non-radioactive if they perturb images in a space that the downstream DM does not learn~\cite{arewmradioactive}. 
In contrast, semantic watermarks aim to introduce a shift in the semantic content of watermarked images~\cite{wen2023treerings,ci2024ringidrethinkingtreeringwatermarking,zhao2023recipe}. Such watermarks can be radioactive: DMs are trained to capture semantic structures in the data distribution, and therefore are likely to learn and propagate the distribution shift induced by a semantic watermark.
However, since these methods do not constrain semantic perturbations, they may substantially modify the image, damaging the fidelity.

We thus propose \tool---the first \emph{semantic-latent} watermarking scheme that is robust, quality-preserving, multi-bit, and radioactive. 
\tool{} achieves radioactivity by embedding a watermark perturbation in the semantic-latent space, also known as the \hspc, and achieving the other desired properties by optimizing the perturbation against robustness, fidelity and multi-bit capacity. Embedding the watermark in h-space allows \tool{} to modify images in a very controlled way to meet these other desired properties, while still maintaining radioactivity \textit{by construction}.
In contrast, existing semantic watermarking methods \cite{wen2023treerings, ci2024ringidrethinkingtreeringwatermarking} manipulate semantic features in a sub-optimal way, resulting in lower watermark imperceptibility without achieving radioactivity~\cite{arewmradioactive}. 
\tool{} is able to optimize its watermark perturbation because we assume that the watermark user can train the perturbation against a representative distribution of images. 
We observe that this condition is generally met, as the watermark user is usually the content owner.

Semantic-latent operation enables effective training, which balances two opposing objectives---minimization of perceived changes to the protected image and radioactivity resulting from these semantic changes.
Remarkably, \tool achieves radioactivity while introducing only minimally perceptible changes, such as slightly adjusting the age of an individual or subtly altering the individual's makeup.

In summary, this paper makes the following contributions:
\begin{myitemize}
    \item We present the novel design of the first semantic-latent space watermark, \tool, which is highly radioactive, while still providing high fidelity, robustness, and multi-bit capacity.
    \item We evaluate \tool under diverse distortions, finetuning conditions, and secret-bit configurations to assess its resistance to perturbations, fidelity, and radioactivity. 
    \item \tool achieves near-perfect watermark detection accuracy, over 95\% bit-level secret recovery accuracy, 1.0 AUC, outperforming the state-of-the-art watermarking scheme.
\end{myitemize}
Upon publication, we will make the source code and datasets for \tool{} openly available.

\section{Background} %
\newpar{Diffusion Models} Diffusion models (DMs) are a class of generative models that have achieved remarkable performance in generating high-quality images, videos, and other modalities. Inspired by concepts from non-equilibrium thermodynamics, DMs consist of two complementary stochastic processes: a \textit{forward diffusion} process, which gradually corrupts input data by adding Gaussian noise until it becomes approximately isotropic Gaussian noise, and a \textit{reverse denoising} process, in which a neural network learns to invert this noising procedure and recover the original data distribution step by step~\cite{ddpm,song2022denoisingdiffusionimplicitmodels,song2019score,song2021score}.
One of the most prevalent and fundamental DMs, Denoising Diffusion Probabilistic Model (DDPM)~\cite{ddpm}, formulates diffusion as a discrete-time Markov chain, adding noise during the forward pass and iteratively denoising from pure noise back to structured data.

\newpar{Interpretable Semantic-latent Space}
Kwon et al.~\cite{kwon2023diffusionmodelssemanticlatent} introduced \textit{semantic-latent space} (or \textit{\hspc}) of DMs, defined as the collection of UNet~\cite{ronneberger2015unet} bottleneck feature maps across timesteps during the reverse diffusion process. 
The UNet bottleneck feature maps are the deepest layers of the UNet for each timestep in the reverse denoising process.
By design, these feature maps, $h_t, t \in T..1$, have the lowest spatial dimensions/resolutions and the most channels/high-level semantics. Therefore, \hspc is the best place to encode compact and semantically meaningful information and enable semantic edits. H-space has three key properties: (1) semantic directions generalize across samples, (2) direction magnitude controls edit strength, and (3) multiple directions can be linearly combined for multi-attribute manipulation. 
Therefore, \hspc enables controlled editing of pre-trained DMs in an interpretable manner. For instance, 
Haas et al.~\cite{haas2024discoveringinterpretabledirectionssemantic} identified semantic directions in h-space using Jacobian analysis and PCA, while Park et al.~\cite{park2023spectral} revealed dominant semantics via spectral Jacobian analysis.

\begin{figure*}[t]
    \centering
    \includegraphics[width=0.85\linewidth]{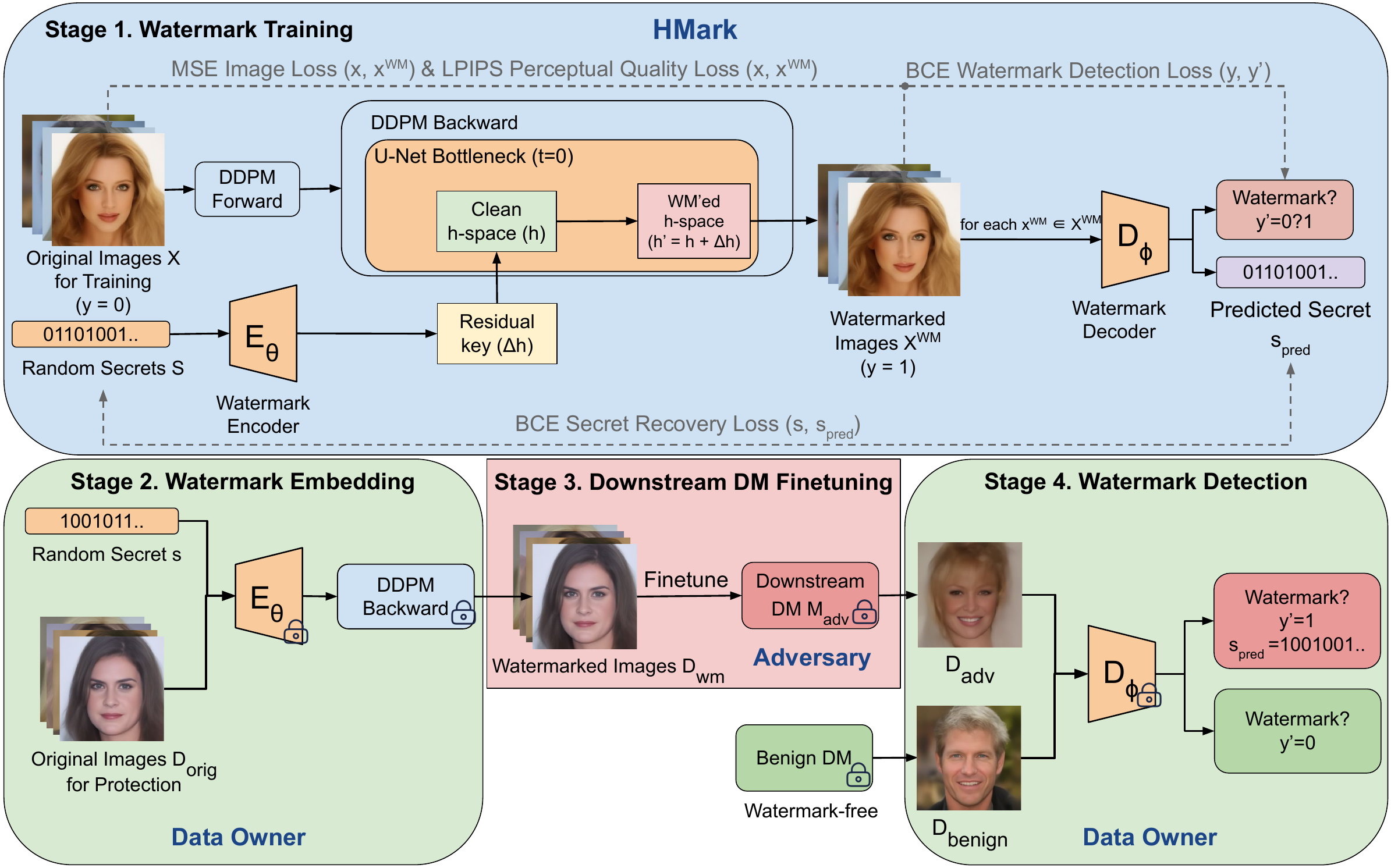}
    \caption{\tool's pipeline: (1) \tool trains a watermarking encoder--decoder on a set of images X and random secrets; (2) Data owner utilizes \tool to embed watermarks on their proprietary data \dorig for IP protection before publishing; (3) The adversary finetunes \madv on data owner's watermark-protected \dwm scraped online; (4) Data owner utilizes \tool to detect potential IP infringement in \dadv. 
    }
    \label{fig:flowchart}
\end{figure*}
\section{Design of \tool}\label{sec:design}

\subsection{Threat Model}\label{subsec:threatmodel}

We consider a scenario in which a data owner seeks to protect their proprietary image dataset \dorig from unauthorized use in DM finetuning. Our threat model contains three entities: the watermarking technology provider (us), the image dataset owner, and an adversary.
To enable copyright protection, the data owner generates the watermarked version of the dataset \dwm by embedding imperceptible watermarks into their proprietary images \dorig.
The adversary controls a DM \madv, obtains an unauthorized copy of the watermarked dataset \dwm, and subsequently finetunes their model through incremental training on \dwm and generates images \dadv.
The data owner aims to determine whether the suspect model \madv has been finetuned on \dwm. The dataset owner is assumed to have closed-box (API query only) access to the suspect model, allowing them to obtain generated samples \dadv but not to inspect the \madv's internal parameters.
\subsection{Design Overview}\label{subsec:overview}
We present \tool, a novel semantic-latent \hspc watermarking scheme. 
Unlike other types of watermarks, our method trains an autoencoder to discover adaptive semantic embedding that effectively balances the trade-off between watermark imperceptibility and detection robustness. 
Given an image $\mathbf{x} \in \mathbb{R}^{H \times W \times 3}$ and a binary secret watermarking message $\mathbf{s} \in \{0,1\}^{B}$ where $B$ denotes the secret length, we formulate the watermarking embedding and detection problem as a constrained optimization task.  The primary objective involves \emph{embedding} $\mathbf{s}$ into $\mathbf{x}$ via a trained \emph{\hspc encoder} to produce a watermarked image $\mathbf{x}_\text{wm}$ while minimizing the perceptual distance $\|\mathbf{x}_\text{wm} - \mathbf{x}\|^2_2$. Simultaneously, we must train a robust \emph{decoder} that is capable of identifying the presence of the watermark and accurately recovering the embedded secret bits. 
The scheme additionally requires that the watermark signal remains detectable after exposure to common image transformations and, critically, survives the \madv finetuning processes (\ie, radioactive).

Our methodology, as shown in Figure~\ref{fig:flowchart}, implements a systematic four-stage pipeline designed to train \tool and assess whether its watermarks propagate from the unauthorized finetuning data into the learned data distribution, thereby contaminating subsequent samples with detectable watermark traces. 

\subsection{\textbf{\tool}'s Training}
\label{sec:train}
The training process involves the watermark encoder $E_\theta$ and decoder $D_\phi$, as outlined in the high-level Algorithm~\ref{alg:training}, Appendix~\ref{app:algo}. In the training process, we apply a custom loss function and distortion-aware input perturbations.

\newpar{H-Space Encoder}
The encoder $E_\theta$ establishes a learnable transformation that maps discrete binary secrets directly to continuous \hspc residuals. 
The watermark embedding procedure leverages the semantic-preserving properties of the UNet bottleneck layer (\hspc) in the reverse diffusion process of DMs.
At the final timestep (\ie, $t=0$) of the reverse diffusion process, the watermark is injected at the UNet bottleneck layer.
\begin{align*}
\mathbf{z}_t^{\text{down}} &= \text{DownBlocks}(\mathbf{x}_t, t), \\
\mathbf{h}_t &= \text{MidBlock}(\mathbf{z}_t^{\text{down}}, t)
\quad \text{(bottleneck features)}, \\
\mathbf{h}_t^{\text{wm}} &= \mathbf{h}_t + \Delta_\mathbf{h_t}
\quad \text{(additive watermark injection)}, \\
\mathbf{x}_{t-1} &= \text{UpBlocks}(\mathbf{h}_t^{\text{wm}}, t).
\end{align*}
The encoder $E_\theta$ is trained to generate residuals $\Delta_\mathbf{h}$ that encode the secret $\mathbf{s} \in \{0,1\}^B$ while minimizing perceptual distortion. $\Delta_\mathbf{h}$ is injected into the last timestep of the reverse diffusion and produces watermarked images $\mathbf{x}_0^{\text{wm}}$.

\newpar{Pixel-Space Decoder}
The detection operates on the image's pixel space. The decoder identifies the presence of watermarks and recovers the embedded secret message: $p_{\text{wm}}, \mathbf{s}_{\text{pred}} = D_\phi(\mathbf{x}_\text{wm})$ where $p_{\text{wm}} \in [0,1]$ is the watermark presence probability 
(\ie, watermark detected if  $p_{\text{wm}} > 0.5$)
, and $\mathbf{s}_{\text{pred}} \in \{0,1\}^B$ is the recovered secret message. $D_\phi$ incorporates a shared convolutional neural network that learns hierarchical representations optimized for both tasks. This shared representation feeds into task-specific heads, where the detection head performs binary classification and the secret extraction head conducts multi-bit regression.

\newpar{Multi-Objective Loss Function}
Our training method employs a multi-objective loss function that addresses the inherent trade-offs between watermark detectability, secret recoverability, and image quality preservation. The multi-component loss function is formulated as:
$\mathcal{L}_{\text{total}} = \lambda_{\text{wm}}\mathcal{L}_{\text{wm}} 
+ \lambda_{\text{secret}}\mathcal{L}_{\text{secret}} + \lambda_{\text{image}}\mathcal{L}_{\text{image}} 
+ \lambda_{\text{lpips}}\mathcal{L}_{\text{lpips}}$.

The \emph{watermark detection} loss $\mathcal{L}_{\text{wm}}$ employs binary cross-entropy (BCE) with logits:
\begin{equation*}
\mathcal{L}_{\text{wm}} = -\frac{1}{N}\sum_{i=1}^{N}\left[y_i \log\sigma(z_i) + (1 - y_i)\log(1 - \sigma(z_i))\right],
\end{equation*}
where $z_i \in \mathbb{R}$ is the decoder's logit output and $y_i \in \{0,1\}$ indicates watermark presence. It optimizes the decoder's ability to distinguish between watermarked and original images, ensuring reliable watermark presence identification.

The \emph{secret recovery} loss $\mathcal{L}_{\text{secret}}$ utilizes bit-wise binary cross-entropy to maximize the accuracy of individual bit predictions, improving reconstruction of the embedded secret message. It is computed over watermarked images ($y_i = 1$):
\begin{align*}
\mathcal{L}_{\text{secret}} 
&= -\frac{1}{|\mathcal{W}| \cdot B}
   \sum_{i \in \mathcal{W}}\sum_{j=1}^{B} \bigg[
      s_{ij}\log\sigma(z_{ij}) \\
&\qquad\qquad\quad
      + (1 - s_{ij})\log(1 - \sigma(z_{ij}))
   \bigg],
\end{align*}
where $B$ is the number of bits, $s_{ij} \in 0,1$ is the $j$-th bit of the $i$-th secret, and
$z_{ij} \in \mathbb{R}$ is the predicted logit.

The \emph{image fidelity} loss $\mathcal{L}_{\text{image}}$ incorporates mean squared error (MSE) in the pixel space:
\begin{equation*}
\mathcal{L}_{\text{image}} = \frac{1}{NCHW}
\cdot \sum_{i=1}^{N}\sum_{c=1}^{C}\sum_{h=1}^{H}\sum_{w=1}^{W}\left(X^{\text{wm}}_{i,c,h,w} - X_{i,c,h,w}\right)^2,
\end{equation*}
where $X^{\text{wm}}$ and $X$ denote the watermarked and original images, respectively. 

Finally, the \emph{perceptual quality} loss component $\mathcal{L}_{\text{lpips}}$ leverages LPIPS~\cite{zhang2018lpips} with normalized inputs to preserve high-level visual characteristics and semantic content:
\begin{equation*}
\mathcal{L}_{\text{lpips}} 
= \frac{1}{N} \sum_{i=1}^{N} 
\text{LPIPS}\!\left(X_i,\; X^{\text{wm}}_i\right),
\quad X \in [-1, 1].
\end{equation*}

It is important to note that radioactivity is not explicitly included in the loss objective. It is achieved by construction rather than through optimization.

\newpar{Distortion-Aware Training}%
We apply a diverse set of distortions during the training process, following SleeperMark's selection~\cite{wang2025sleepermarkrobustwatermarkfinetuning}, which includes additive Gaussian noise, spatial filtering through Gaussian blur convolutions, compression, photometric variations through brightness and contrast adjustments, and geometric transformations via multi-scale resizing.
This comprehensive distortion regime operates with a uniform probability 
throughout training, ensuring that the decoder develops invariance to these perturbations without overfitting to any transformation.

\subsection{Theoretical Intuition on Radioactivity}\label{subsec:theory}

\tool injects watermark information into the UNet bottleneck (\hspc), which results in semantic shifts, rather than direct pixel modifications. When a watermark residual $\Delta_{\mathbf{h}}$ is added in the latent space, all images that embed the same secret receive the same latent shift, even though each image has different pixel-level content. Because the watermark perturbation is constant in \hspc, the effect is coherent throughout the dataset. That is, \tool introduces a shift to the distribution rather than isolated perturbations to individual samples. 
Diffusion models learn a score function~\cite{ddpm}, the gradient of the log-density $\nabla_{\mathbf{x}}\log p(\mathbf{x})$, which characterizes the data distribution. Thus, a coherent distribution shift in the training data can be reflected in the learned score function. Empirically, we observe that this allows watermark information to persist after finetuning.

Formally, let $P_{\text{orig}}$ be the original data distribution. Injecting a latent residual modifies the last UNet in the reverse diffusion process as follows, producing the watermarked dataset distribution:
\[
T_{\Delta_{\mathbf{h}}}(\mathbf{x})
=
\mathrm{UpBlocks}\!\big(\mathbf{h} + \Delta_{\mathbf{h}},\; 0\big)
\]
\[
P_{\text{wm}} = (T_{\Delta_{\mathbf{h}}})_{\#} P_{\text{orig}}.
\]

To understand how the latent residual affects the resulting image, we linearize the decoder around
each image’s latent bottleneck point. The decoder's Jacobian $\mathbf{B}(\mathbf{x})$ maps the change
in latent representation to the pixel space:
\[
\mathbf{x}^{\text{wm}}
\approx
\mathbf{x} + \mathbf{B}(\mathbf{x})\,\Delta_{\mathbf{h}},
\qquad
\mathbf{B}(\mathbf{x})
=
\frac{\partial\,\mathrm{UpBlocks}}{\partial \mathbf{h}}
\Big|_{(\mathbf{h}_{0},\mathbf{z}^{\text{down}}_{0})}.
\]

Even though $\mathbf{B}(\mathbf{x})$ differs for each sample, $\Delta_{\mathbf{h}}$ is the same for all the images
when the same secret is embedded. Aggregated over the dataset, this results in a distribution shift in expectation:
\[
\Delta \boldsymbol{\mu}_x
\coloneqq
\mathbb{E}_{P_{\mathrm{wm}}}[\mathbf{x}]
-
\mathbb{E}_{P_{\text{orig}}}[\mathbf{x}]
\approx
\Big(\mathbb{E}_{\mathbf{x}\sim P_{\text{orig}}}[\mathbf{B}(\mathbf{x})]\Big)\,\Delta_{\mathbf{h}}.
\]
When the downstream model trains on a watermarked dataset \dwm, the model learns the watermarked score function. As a result, the generation of the downstream model converges to the modified distribution, $\tilde P^{\text{gen}} \approx P_{\mathrm{wm}}.$
This provides an intuition as to why watermarks embedded by \tool are radioactive by construction, as they embed their signals into the dataset distribution which is later learned by the downstream model.

\section{Evaluation} \label{sec:eval}
In this section, we discuss experiments on \tool, evaluating its radioactivity, robustness, imperceptibility, and capacity.
All experiments are performed on a machine with two Intel Xeon Gold 6548Y processors and four Nvidia H100 GPUs having 96GB of high-bandwidth memory. 

\newpar{Models and Dataset}
At Stages 1 and 2 (\ie, \tool's training and embedding in Figure~\ref{fig:flowchart}), we use the \hspc-enabled semantic DDPM
~\cite{haas2024discoveringinterpretabledirectionssemantic} for watermark injection.
At Stages 3 and 4, we employ a pixel-space DDPM~\cite{ddpm} as the attacker's \madv. To ensure a closed-box setting, we use a pre-trained model checkpoint \texttt{google/ddpm-celebahq-256} on the CelebA-HQ dataset~\cite{celebahq}.
The finetuning techniques we adapt are the basic DDPM's UNet finetuning and LoRA ($\texttt{rank}=8, 16, 32$)~\cite{hu2021loralowrankadaptationlarge}. We use finetuning scripts from Hugging Face Diffusers' official GitHub repository~\cite{diffusers} with minimal changes to support DDPM. 
Due to space constraints, we present the results on LoRA in this section, and the UNet finetuning results 
are discussed in Appendices~\ref{app:distortion},~\ref{app:sec_size} and~\ref{app:discuss_steps}.

At Stage 1, to train \tool, we use 100{,}000 generated images from the semantic DDPM with random secret bits under different random seeds. 
At Stage 2, we use a protected dataset \dorig (see Section~\ref{subsec:threatmodel}) of 10{,}000 samples to generate a \dwm. Note that \dorig{} is not part of our training dataset, thereby ensuring generalizability.
At Stage 3, the adversary uses \dwm to finetune \madv.
At Stage 4, we evaluate \tool's detector's effectiveness in distinguishing \dadv (\ie, outputs of \madv) from \dbenign (\ie, outputs of a benign downstream DM without finetuning). 
We study the impact of the size of \dwm, \dwmsize, in Section~\ref{subsec:radioactive}.

Although in our threat model, the adversary cannot acquire \dorig (\ie, the same dataset as \dwm but watermark-free) and thus cannot generate \dclean (\ie, outputs of a downstream DM finetuned on \dorig), we still evaluate \tool's effectiveness to distinguish \dadv from \dclean. This evaluation allows us to compute the false positive rate of \tool, and confirm the authenticity of its detection results.

In evaluation, finetuning is 9{,}000-step, which is selected based on examples in Hugging Face Diffusers module~\cite{diffusers, huggingface_lora_diffusers}. We analyze the effect of the number of finetuning steps in Section~\ref{subsec:acc_steps}.
Other training and finetuning hyperparameter choices are presented in Appendix~\ref{app:hyper}. Although the evaluation focuses on embedding 8-bit secrets,
we also study \tool's support of different secret sizes, and results are discussed in Section~\ref{subsec:sec_size}.

\newpar{Evaluation Metrics and Comparison}
We evaluate \tool's effectiveness using watermark detection accuracy, $\text{Acc}_{\text{WM}} = \frac{TP + TN}{TP + TN + FP + FN}$; bit-level secret accuracy, $\text{Acc}_{\text{bit}} \;=\; \frac{1}{B}\sum_{j=1}^{B} \left[\hat{s}_{j} = s_{j}\right]$, where $s_{j}$ is the $j$-th index of the ground-truth secret $\mathbf{s}$ and $\hat{s}_j$ is the $j$-th index of the predicted secret $\hat{\mathbf{s}}$ by the detector, $\mathbf{s},\,\hat{\mathbf{s}} \in \{0,1\}^B$; precision, $P = \frac{TP}{TP + FP}$; recall rate, $R = \frac{TP}{TP + FN}$; and F1 score, $F_1 = 2 \times \frac{P \times R}{P + R}$. We also use Receiver Operating Characteristic (ROC) and Area Under the Curve (AUC)~\cite{AUCROC}.
We quantitatively measure visual fidelity using Mean Squared Error (MSE)~\cite{mse_statbook}, Structural Similarity Index (SSIM)~\cite{wang2004ssim}, Learned Perceptual Image Patch Similarity (LPIPS) with AlexNet~\cite{zhang2018lpips}, and Fréchet Inception Distance (FID)~\cite{heusel2017fid}. %

\begin{table}[ht!]
\centering
\footnotesize
\setlength{\tabcolsep}{1.25pt}
\caption{
Evaluation of \tool’s effectiveness, compared to Recipe watermarking scheme~\cite{zhao2023recipe}.
Stage 2 results show the immediate post-embedding watermark effectiveness, while Stage 4 results show \tool~'s radioactivity across finetuning methods.
}
\begin{tabular}{lllcccll}
\toprule
\textbf{Stg.} & \textbf{Scheme} & \textbf{Finetune} & \textbf{\wmacc (\%)} & \textbf{\bitacc (\%)} & \textbf{P (\%)} & \textbf{R (\%)} & \textbf{F1 (\%)} \\
\midrule
\multirow{2}{*}{2} & Recipe & NA & 99.3 & 57.2 & 98.6 & 100.0 & 99.3 \\
                   & \tool & NA & 99.0 & 97.4 & 98.1 & 100.0 & 99.1 \\
\midrule
\multirow{4}{*}{4} & \multirow{2}{*}{Recipe} & UNet & 50.8 & 16.7 & 72.4 & 2.7 & 5.2 \\
                   &                         & LoRA & 50.6 & 16.1 & 77.2 & 2.0 & 3.9 \\
                   & \multirow{2}{*}{\tool}  & UNet & 98.1 & 87.4 & 97.3 & 99.0 & 98.2 \\
                   &                         & LoRA & 98.6 & 95.1 & 97.2 & 100.0 & 98.6 \\
\bottomrule
\end{tabular}
\label{tab:tool_eval}
\end{table}

\begin{figure}[ht!]
\centering
\includegraphics[width=0.85\linewidth]{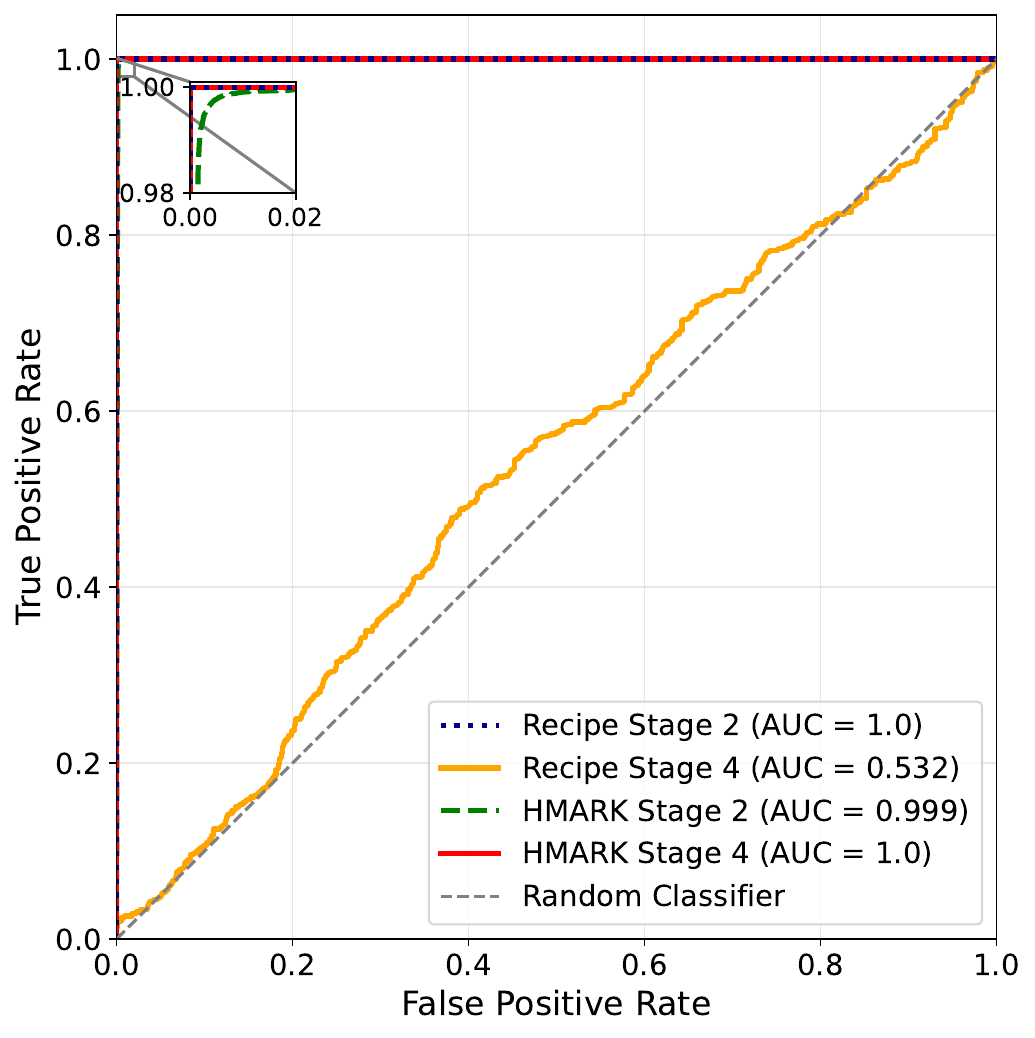}
\caption{
ROC curves of Recipe versus \tool at Stage 2 and Stage 4 when \madv is LoRA finetuned.
}
\label{fig:roc}
\end{figure}

\begin{figure}[ht!]
\centering
\includegraphics[width=0.9\linewidth]{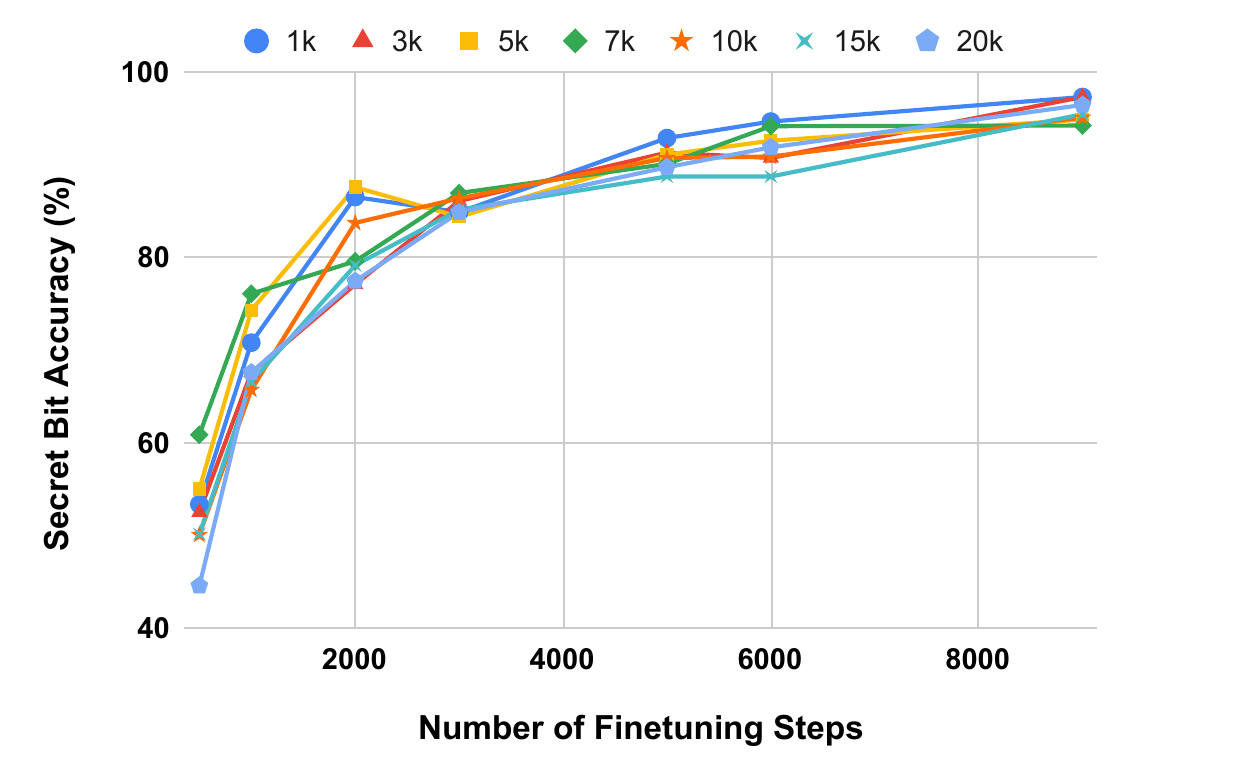}
\caption{
Impact of \dwmsize on \bitacc and the evolution of \tool's performance as the number of finetuning steps increases when \madv is LoRA finetuned.
}
\label{fig:trend}
\end{figure}

\subsection{Effectiveness and Radioactivity} \label{subsec:radioactive}
To evaluate \tool's effectiveness, we collect the results from two detection points: (1) watermarked images (\dwm) generated from stage 2, and (2)  generated outputs (\dadv) from the adversary's \madv at stage 4. 
We compare \tool with a well-known watermarking scheme \textit{Recipe}~\cite{zhao2023recipe}.  We select Recipe as it is the only recent DDPM-supporting watermark we are aware of at the time of writing.

Table~\ref{tab:tool_eval} presents Recipe and \tool's watermark performances when 8-bit secrets are embedded.
Tested on \dwm and \dorig, both \tool and Recipe achieve high \wmacc of around 99\%, but \tool excels at secret recovery with \bitacc of 97.4\% versus Recipe's 57.2\%. 

After \madv finetuning, evaluated on 100 samples from \dadv and 100 samples from \dbenign, \wmacc still remains near-perfect for \tool, demonstrating the radioactivity of the embedded watermark. The consistently high \bitacc (87.43\% with UNet finetuning and 95.07\% with LoRA $\texttt{rank}=32$) and F1~scores further confirm \tool's robustness and reliability across adversarial settings. 
We also study the effect of LoRA ranks and discuss the results of lower ranks (16 and 8) in Appendix~\ref{app:lora_rank}. We observe that \tool's \bitacc can exceed 95\% for all the ranks.

In contrast, Recipe's \wmacc drops to almost 50\% for both finetuning techniques.
Based on the ROC analysis in Figure~\ref{fig:roc}, we observe that at Stage 2, both schemes demonstrate excellent \wmacc with AUC values of 0.999 and 1.0, respectively. At Stage 4, however, \tool maintains its reliable detection capability with an AUC of 1.0, whereas Recipe exhibits substantial degradation (\ie, AUC drops to 0.532), barely exceeding random guessing. 

To further verify \tool's detection authenticity and robustness against false attribution, we evaluate its behavior on 200 generated outputs \dclean that were not watermarked. \tool achieves 93.93\% \wmacc with 6.07\% FPR and 30.27\% \bitacc. These results demonstrate that \tool reliably identifies almost all outputs of \madv as unwatermarked, even though the underlying training images are identical to those of \dwm but without watermarks. Although a small fraction of benign samples are mistakenly flagged as watermarked, the 30.27\% \bitacc confirms that \tool cannot extract any valid secret bits from them, thereby validating the watermark's detection authenticity. Detailed results for different distortions are in Table~\ref{tab:original_image_finetuning} in Appendix~\ref{app:additional_results}.
Detailed evaluation results for Recipe are presented in Appendix~\ref{app:recipe}.

As shown in Figure~\ref{fig:trend}, \tool's effectiveness is relatively insensitive to the number of watermarked images \dwmsize. \tool is highly reliable in watermark detection even when \dwmsize is as \textit{small}~\cite{small} as 1{,}000, with \wmacc constantly around 98.5\% for all \dwm sizes. 
Figure~\ref{fig:trend} also confirms the theoretical intuition (Section~\ref{subsec:theory})---\bitacc improves, as the model becomes better at understanding the finetuning distribution with each step.

Overall, the results indicate that \tool is radioactive and reliable in watermark detection.
Detailed results are presented in Table~\ref{tab:num_training_steps} in Appendix~\ref{app:training_steps}.

\begin{table}[ht!]
\centering
\setlength{\tabcolsep}{3pt}
\footnotesize
\caption{
Performance of \tool{} evaluated on \madv's generated outputs (\dadv) at Stage 4 under different distortions. \madv is trained with LoRA finetuning script ($\texttt{rank} = 32$). 
}
\begin{tabular}{lccccc}
\toprule
\textbf{Distortion} & \textbf{\wmacc (\%)} & \textbf{\bitacc (\%)} & \textbf{P (\%)} & \textbf{R (\%)} & \textbf{F1 (\%)} \\
\midrule
Identity        & 99.00 & 94.00 & 98.04 & 100.00 & 99.01 \\
Gaussian Noise  & 99.00 & 95.75 & 98.04 & 100.00 & 99.01 \\
Brightness      & 98.50 & 94.38 & 97.09 & 100.00 & 98.52 \\
Contrast        & 99.00 & 94.50 & 98.04 & 100.00 & 99.01 \\
Blur            & 98.50 & 95.50 & 97.09 & 100.00 & 98.52 \\
JPEG            & 99.00 & 95.25 & 98.04 & 100.00 & 99.01 \\
Resize          & 97.00 & 97.12 & 94.34 & 100.00 & 97.09 \\
\midrule
\textbf{Average} & \textbf{98.57} & \textbf{95.07} & \textbf{97.22} & \textbf{100.00} & \textbf{98.59} \\
\bottomrule
\end{tabular}
\label{tab:distortion_8_lora_32}
\end{table}

\subsection{Robustness} \label{subsec:robust}
To evaluate \tool's resistance to common image distortions,
we simulate a range of realistic perturbations, whose parameter settings are provided in Appendix~\ref{app:distortion}. These distortions emulate both intentional post-processing edits and natural image degradations that may occur during acquisition, compression, or transmission in real-world scenarios. Specifically, we consider six types of perturbations: Gaussian noise addition, brightness adjustment, contrast variation, blurring, JPEG compression, and resizing~\cite{wang2025sleepermarkrobustwatermarkfinetuning}.

Table~\ref{tab:distortion_8_lora_32} demonstrates that \tool consistently achieves near-perfect \wmacc, high \bitacc, and a 100\% recall rate, regardless of common perturbations
applied to the outputs of \madv.
These results confirm \tool's radioactivity. %

Stage 2 results (high \wmacc and \bitacc) are shown in Table~\ref{tab:distortion} in Appendix~\ref{app:distortion}.
Comparative analyses of \tool across different finetuning scripts and secret bit lengths are provided in Appendices~\ref{app:distortion} and~\ref{app:sec_size}.

\begin{table}[ht!]
\centering
\setlength{\tabcolsep}{2pt}
\footnotesize
\caption{Fidelity metrics for different secret bit configurations collected on 10{,}000 (clean, watermarked) image pairs. ($\pm$ represents the standard deviation). Lower is better for MSE, LPIPS, and FID; higher is better for SSIM. Result for Recipe is tested on 8-bit secret.}
\begin{tabular}{lllll}
\toprule
\textbf{\# Bits} & \textbf{MSE} ($\downarrow$) & \textbf{SSIM} ($\uparrow$) & \textbf{LPIPS} ($\downarrow$) & \textbf{FID} ($\downarrow$) \\
\midrule
Recipe & 2.47e-4 $\pm$ 3.5e-5 & 0.9747 $\pm$ 0.0035  & 0.141 $\pm$ 0.0417 & 30.4025 \\
\midrule
8  & 1.49e-4 $\pm$ 6.8e-5 & 0.9725 $\pm$ 0.0046 & 0.0511 $\pm$ 0.0155 & 8.3161 \\
16 & 1.72e-4 $\pm$ 9.1e-5 & 0.9683 $\pm$ 0.0056 & 0.0605 $\pm$ 0.0182 & 7.6229 \\
32 & 1.38e-4 $\pm$ 9.6e-5 & 0.9740 $\pm$ 0.0044 & 0.0472 $\pm$ 0.0149 & 6.8820 \\
\bottomrule
\end{tabular}
\label{tab:imperceptible}
\end{table}

\subsection{Fidelity} \label{subsec:invisible}

We evaluate watermark fidelity and invisibility in Table~\ref{tab:imperceptible}. Overall, images with watermarks \dwm exhibit high perceptual quality and imperceptibility. Representative examples are shown in Figure~\ref{fig:orig_wm_pairs} and Figure~\ref{fig:appendix_orig_wm_grid} in Appendix~\ref{app:invisible}. 

The pixel-level distortion, measured by MSE, remains low (on the order of $10^{-4}$) with small variance across secret bit lengths, indicating negligible embedding artifacts.
SSIM measures the structural similarity by considering, among other metrics, luminance, contrast, and structural components. 
\dwm's SSIM values are over 96\%, indicating that the structural and perceptual content of the images remains highly consistent after watermarking. 
LPIPS leverages deep neural network features to measure perceptual differences between image pairs (\ie, 0 indicates identical perception). In our case, we obtain favorable LPIPS scores (0.04--0.06), indicating that \dwm are perceptually indistinguishable from \dorig. 
Lastly, FID evaluates the distributional similarity between real and generated (in our case, \dorig and \dwm). Our FID scores (6.9--8.3) demonstrate that the overall distribution and fidelity of the images are well preserved. 
Compared to Recipe, \tool's watermarked image fidelity is comparable in SSIM and substantially better in other metrics.

Notably, these metrics are consistent across 8-, 16-, and 32-bit secrets, suggesting that watermark imperceptibility is insensitive to embedding capacity.
Figure~\ref{fig:generated_steps_lora32_compare} in Appendix~\ref{app:semantic_shift} presents images generated from \madv finetuned on \dorig, \dwm with 8-bit secret, \dwm with 16-bit secret, and \dwm with 32-bit secret. 
As finetuning progresses, outputs appear more natural, making it harder for an adversary to distinguish clean from watermarked images.

\begin{table}[ht!]
\centering
\footnotesize
\caption{
\tool{}'s performance on radioactivity to \madv finetuned with LoRA ($\texttt{rank}=32$) with different secret bit lengths.
}
\begin{tabular}{lccccc}
\toprule
\textbf{\# Bits} & \textbf{\wmacc (\%)} & \textbf{\bitacc (\%)} & \textbf{P (\%)} & \textbf{R (\%)} & \textbf{F1 (\%)} \\
\midrule
8   & 98.57 & 95.07 & 97.22 & 100.00 & 98.59 \\
16  & 99.50 & 93.73 & 99.43 & 99.14 & 99.50 \\
32  & 90.71 & 68.70 & 87.88 & 94.57 & 91.04 \\
\bottomrule
\end{tabular}
\label{tab:bits_lora}
\end{table}

\subsection{Secret Size Analysis}\label{subsec:sec_size}
Supporting increasing secret sizes is important as it allows assigning exponentially more unique identifiers to watermark users.
Table~\ref{tab:bits_lora} shows \tool's performance averaged across various distortions when different secret sizes (8, 16, 32 bits) are embedded and transmitted during downstream DDPM finetuning with LoRA. 

Following our theoretical intuition (Section~\ref{subsec:theory}), as more information is communicated through the distribution shift with \tool, it becomes harder for the downstream DM to learn this shift. Thus, the recovery degrades as the secret size grows with the same number of finetuning steps.
Results in Table~\ref{tab:bits_lora} confirm this theoretical intuition.
\tool achieves 95.07\% and 93.73\% \bitacc when the secret size is 8 and 16 bits, respectively, and, as we double the secret size to 32, \bitacc drops to 68.70\%. %
To improve these results, even when the number of finetuning steps is limited, 
\tool supports Bose-Chaudhuri-Hocquenghem (BCH) error correction code~\cite{BoseRayChaudhuri1960}, which helps \tool to achieve \bitacc of above 90\%. See details in Appendix~\ref{app:bch}.

Notably, with growing secret sizes, \wmacc decreases only gradually (from 99.50\% to 90.71\%), precision and recall remain high, showing that \tool maintains reliable watermark detectability even at larger bit lengths.
Detailed tabularized results on LoRA and default UNet finetuning are presented in Appendix~\ref{app:sec_size}.

\subsection{Relationship of \textbf{\bitacc} and Finetuning}\label{subsec:acc_steps}

For most part of the evaluation, we set the finetuning step count to 9{,}000, following Hugging Face Diffusers’ GitHub sample configurations~\cite{diffusers}.
However, we also analyze how \tool{}’s performance evolves during finetuning, recording evaluation metrics 
at several step configurations, ranging from 500 to 30{,}000 steps.

When \madv{} is trained using LoRA (\ie, for $\texttt{rank}=8, 16, 32$), we observe that higher ranks yield strong watermark detection early on. For example, with $\texttt{rank}=32$, \wmacc{} starts at 94.21\% at step 500, surpasses 96\% soon after, and exceeds 98\% beyond 4{,}000 steps. Meanwhile, \bitacc{} improves more gradually—remaining under 80\% before 2{,}000 steps, crossing 90\% around 4{,}000 steps, and reaching over 95\% by 9{,}000 steps.
This trend suggests that \tool{} maintains high \wmacc regardless of the duration of \madv's finetuning. 
Additional training mainly boosts \bitacc, but the improvement quickly saturates at a relatively modest level, indicating diminishing returns beyond approximately 6k--9k steps.
A similar trend is observed when \madv is finetuned with default UNet script. Here, \wmacc{} stays above 97\% throughout, confirming stable watermark detectability, while \bitacc{} gradually rises before leveling off after 6{,}000 steps.
Specifically, \tool{} achieves around 61\% \bitacc{} within the first 3{,}000 steps, increases to 81.86\% after 5{,}000 steps, and surpasses 85\% beyond 6{,}000 steps, eventually reaching 87.64\% at 30{,}000 steps. 
Tabularized results are presented in Tables \ref{tab:finetune_steps} and \ref{tab:finetune_steps_lora_all} in Appendix~\ref{app:discuss_steps}.

\section{Limitations} \label{sec:discuss}

Although \tool achieves radioactivity agnostically to the finetuning method (LoRA and default UNet finetuning)~\cite{hu2021loralowrankadaptationlarge,diffusers}, finetuning algorithms with latent-space diffusion process such as Latent Diffusion Models (LDMs)~\cite{rombach2022ldm} may require additional training. While we acknowledge this limitation, it is orthogonal to our main contribution, which is a multi-bit semantic-latent space watermarking scheme that is high-fidelity, robust and radioactive by construction.

\section{Related Work} \label{sec:related}
\newpar{Watermarking} Watermarking schemes are widely used for IP protection and ownership verification. 
Watermarks have been injected into Convolutional Neural Networks using feature-domain patterns~\cite{LoneNeuron}, colour-space triggers~\cite{colorbackdoor}, and architectural backdoors~\cite{langford2024architecturalneuralbackdoorsprinciples}. While effective for classification tasks with discrete outputs, these methods struggle in generative settings, where outputs are continuous and high-dimensional, making consistent trigger responses unreliable.
Generative adaptations such as~\cite{shao2024explanationwatermarkharmlessmultibit} show promise but remain vulnerable to removal attacks~\cite{hu2024weaknessesbackdoorbasedmodelwatermarking}. Latent-space watermarking approaches, such as WaDiff~\cite{wadiff}, PRC~\cite{gunn2025undetectablewatermarkgenerativeimage} and Stable Signature~\cite{stable_signature}, embed watermarks directly into the generative process. However, these methods still face the challenge of achieving radioactivity~\cite{arewmradioactive}, ensuring all outputs from the downstream model carry the watermark.
Invisible pixel-space watermarks~\cite{dwt_dct, stegastamp, rivagan} are provably removable using generative AI~\cite{zhao2024invisibleimagewatermarksprovably}, motivating semantic watermarks that modify high-level content while preserving meaning. Examples include Tree-Ring~\cite{wen2023treerings}, which embeds patterns in the initial noise vector, and RingID~\cite{ci2024ringidrethinkingtreeringwatermarking}, which extends this idea to multi-bit capacity. 
SleeperMark~\cite{wang2025sleepermarkrobustwatermarkfinetuning} takes a different approach, disentangling watermark information from semantics to survive DM finetuning. As for \cite{meintz2025radioactivewatermarksdiffusionautoregressive}, although it is radioactive, its watermarks are only zero-bit.

\newpar{Cloaking} Cloaking schemes share a similar goal with ours, offering an alternative means of protecting datasets and preventing unauthorized model training through adversarial redirection. Glaze~\cite{shan2025glazeprotectingartistsstyle} applies imperceptible  ``style cloaks'' that mislead models attempting to imitate an artist's style. MetaCloak~\cite{metacloak} uses meta-learning to generate transferable perturbations robust to transformations, while Anti-Dreambooth~\cite{vanle2023antidreamboothprotectinguserspersonalized} disrupts personalized text-to-image finetuning via targeted noise. NightShade~\cite{shan2024nightshadepromptspecificpoisoningattacks} further introduces perturbations that both cloak and poison, causing DMs trained on protected data to yield corrupted outputs. Despite their promise, cloaking methods face key limitations. They require proactive protection of every image, and their perturbations can be neutralized by standard pre-processing or robust training~\cite{foersterlightshed}. Once disclosed, the perturbation algorithm allows for removal and indefinite reuse of the original data, rendering protection ineffective~\cite{foersterlightshed}. Moreover, cloaking offers no mechanism for ownership verification or proof of unauthorized use. In contrast, our approach develops a robust, radioactive watermarking scheme operating in the semantic-latent space of DMs to enable verifiable dataset ownership and reliable detection of misuse.

\newpar{Data Poisoning}
Data poisoning is a related yet orthogonal research direction to our work. Poisoning attacks aim to maliciously manipulate training data to alter model behavior, degrade performance, or implant backdoors, hence, compromising model integrity and reliability~\cite{zhao2025survey}. Recent works highlight the diversity of poisoning strategies and threat models, from label flip and feature injection to stealthy clean-label poisoning~\cite{DeGaspari2023WildPatterns,nguyen2024manipulatingrecommendersystemssurvey}.
Modern approaches in generative model data poisoning can perturb diffusion and image-generation pipelines through bilateral effects or optimization-based attacks~\cite{he2025sharpnessawaredatapoisoningattack,Sun_2023,pan2024trojanhorsescastlewalls}. Although a watermarking scheme has a similar technical idea to data poisoning, it serves a protective role by embedding imperceptible yet verifiable signatures into data to establish ownership or trace unauthorized use, without adversarially degrading model's utility.

\section{Conclusion}
The rapid innovations of generative DMs have amplified the tension between human creativity, IP rights, and generative AI technologies. Existing watermarking schemes fall short in reliably identifying unauthorized model finetuning on proprietary datasets. 
To address this critical gap, we introduce \tool, a novel multi-bit semantic-latent watermarking scheme. \tool's watermarks are radioactive, imperceptible, multi-bit, and robust. By embedding semantically meaningful and persistent signals within finetuning data, \tool enables transparent ownership verification on generated outputs of the downstream models, exceeding 98\% watermark detection accuracy, 95\% secret bit recovery rate, and perfect AUC, while preserving visual fidelity. Comprehensive evaluation demonstrates \tool's effectiveness under diverse distortions and finetuning settings. Together, these advances represent a crucial step towards responsible and legally compliant generative AI development.

{
\small
\bibliographystyle{ieeenat_fullname}
\bibliography{ref}
}

\clearpage
\appendix

\phantomsection
\addcontentsline{toc}{part}{Appendix} %
\section*{Appendix}
\newpar{Overview} In this supplementary material, we present the following:

\begingroup
  \etocsetnexttocdepth{2}
  \etocsettocstyle{}{}      %
  \localtableofcontents
\endgroup

\begin{algorithm} 
\caption{Stage 1 \tool Training}\label{alg:training}
\begin{algorithmic}[1]
\Require Training dataset $\mathcal{D}$, encoder $E_\theta$, detector $D_\phi$
\Require Loss weights $\lambda_{\text{wm}}$, $\lambda_{\text{secret}}$, $\lambda_{\text{image}}$, $\lambda_{\text{lpips}}$
\Ensure Trained encoder $E_\theta^*$, detector $D_\phi^*$
\State Initialize $\theta$, $\phi$ randomly
\For{epoch $= 1$ to $N_{\text{epochs}}$}
    \For{$\mathbf{x}$ in $\mathcal{D}$}
        \State Sample random secret $\mathbf{s} \sim \{0,1\}^B$
        \State \textit{// Watermark embedding}
        \State $\Delta_h, \mathbf{s}_{\text{self}} \leftarrow E_\theta(\mathbf{s})$
        \State $\mathbf{x}_{wm} \leftarrow \text{embed}(\mathbf{x}, \Delta_h)$
        \State \textit{// Apply random distortion with probability $p_{\text{distort}}$}
        \If{$\text{random}() < p_{\text{distort}}$}
            \State $\mathbf{x}_{wm} \leftarrow \text{distort}(\mathbf{x}_{wm})$
        \EndIf
        \State \textit{// Detection}
        \State $p_w, \mathbf{s}_w \leftarrow D_\phi(\mathbf{x}_{wm})$
        \State $p_c, \mathbf{s}_c \leftarrow D_\phi(\mathbf{x})$
        \State \textit{// Compute losses}
        \State $\mathcal{L}_{\text{wm}} = \text{BCE}(p_w, 1) + \text{BCE}(p_c, 0)$
        \State $\mathcal{L}_{\text{secret}} = \text{BCE}(\sigma(\mathbf{s}_w), \mathbf{s}) + \text{BCE}(\mathbf{s}_{\text{self}}, \mathbf{s})$
        \State $\mathcal{L}_{\text{image}} = \|\mathbf{x}_{wm} - \mathbf{x}\|_2^2$
        \State $\mathcal{L}_{\text{lpips}} = \text{LPIPS}(\mathbf{x}_{wm}, \mathbf{x})$
        \State $\mathcal{L} = \lambda_{\text{wm}}\mathcal{L}_{\text{wm}} + \lambda_{\text{secret}}\mathcal{L}_{\text{secret}} + \lambda_{\text{image}}\mathcal{L}_{\text{image}} + \lambda_{\text{lpips}}\mathcal{L}_{\text{lpips}}$
        \State Update $\theta, \phi$ using $\nabla \mathcal{L}$
    \EndFor
\EndFor
\end{algorithmic}
\end{algorithm}

\begin{algorithm}
\caption{\tool Pipeline}\label{alg:pipeline}
\begin{algorithmic}[1]
\Require Original dataset \dorig owned by dataset owner, suspect DM \madv owned by adversary
\State \textbf{Stage 1:} Train watermark encoder $E_\theta$ and detector $D_\phi$ and produces trained encoder $E_\theta^*$, detector $D_\phi^*$
\State \textbf{Stage 2:} Dataset owner generates watermarked dataset \dwm from \dorig with \tool
    \For{$\mathbf{x}$ in \dorig}
        \State $\mathbf{s} \leftarrow$ fixed secret pattern with length $B$, $\mathbf{s} \in \{0,1\}^{B}$ 
        \State $\mathbf{x}_{wm} \leftarrow \text{embed}(\mathbf{x}, E_\theta^*(\mathbf{s}))$
        \State Save $\mathbf{x}_{wm}$ to \dwm
    \EndFor
\State \textbf{Stage 3:} Adversary finetunes downstream DM \madv
    \State \madv $\leftarrow$ finetune(\madv, \dwm)
\State \textbf{Stage 4:} Dataset owner verifies IP infringement and test radioactivity
    \For{$i = 1$ to $N_{\text{test}}$}
        \State \dadv $\leftarrow$ \madv.generate()
        \State $p_{\text{detect}}, \mathbf{s}_{\text{pred}} \leftarrow D_\phi^*$(\dadv)
        \State Record watermark detection rate and secret accuracy
    \EndFor
\end{algorithmic}
\end{algorithm}

\section{Algorithms}\label{app:algo}
We present \tool's watermark encoder--decoder training algorithm in Algorithm~\ref{alg:training}. \tool's full pipeline is shown in Algorithm~\ref{alg:pipeline}.

\section{Hyperparameter Configurations for Training and Downstream Finetuning}\label{app:hyper}
The hyperparameters at Stage~1 watermark encoder--decoder training are as follows. 
The training is conducted with the number of epochs set to 100, batch size set to 16, and learning rate ($\texttt{lr}$) set to $0.0003$. 
As for the loss function, we set $\lambda_\text{wm} = 5.0$, $\lambda_\text{secret} = 3.0$, $\lambda_\text{image} = 1.5$, and $\lambda_\text{lpips} = 2.0$. 
We set the distortion probability $p_{\text{distort}} = 1.0$. 
Finally, the number of secret bits is set to $8$, $16$, and $32$, respectively, based on the specific experimental configurations.  
The finetuning hyperparameters at Stage~3 are set as follows. 
By default, for both finetuning scripts, we set \texttt{resolution} to 256, \texttt{train\_batch\_size} to 16, \texttt{mixed\_precision} to ``no'', \texttt{checkpointing\_steps} to 500, and \texttt{lr\_warmup\_steps} to 500. 
For LoRA finetuning, we set $\texttt{lr} = 0.0001$ and \texttt{rank} to 8, 16, or 32 based on the specific experiment; and for the default UNet finetuning, we set $\texttt{lr} = 0.0002$ with the \texttt{use\_ema} flag set to ``True''.

\begin{table}[th!]
\centering
\caption{Performance of \tool evaluated on \dclean under various distortions when \madv is finetuned on watermark-free \dorig using LoRA finetuning ($\texttt{rank}=32$).}
\setlength{\tabcolsep}{3pt}

\label{tab:original_image_finetuning}
\begin{tabular}{lccc}
\toprule
\textbf{Distortion} & \textbf{\wmacc (\%)} & \textbf{\bitacc (\%)} & \textbf{FPR (\%)} \\
\midrule
Identity        & 96.0  & 29.44  & 4.0  \\
Gaussian Noise  & 95.5  & 30.94  & 4.5  \\
Brightness      & 96.0  & 30.0  & 4.0  \\
Contrast        & 96.0  & 30.06  & 4.0  \\
Blur            & 93.0  & 30.0  & 7.0  \\
JPEG            & 95.0  & 29.31  & 5.0  \\
Resize          & 86.0  & 32.12  & 14.0 \\
\midrule
\textbf{Average} & \textbf{93.93} & \textbf{30.27} & \textbf{6.07} \\
\bottomrule
\end{tabular}
\end{table}

\section{Additional Results to Section~\ref{subsec:radioactive}}\label{app:additional_results}
Table~\ref{tab:original_image_finetuning} records \tool's watermark detection when \madv is LoRA finetuned on samples from \dorig (\ie,~all of them should be True Negatives). 93.93\% \wmacc means that \tool correctly identifies 93.93\% of the images as non-watermarked and 6.07\% of them are incorrectly identified as watermarked. \bitacc being 30.27\% means that \tool cannot extract the correct majority of the watermark's bits from \madv's generated outputs, which is the expected behavior.

\begin{table}[th!]
\centering
\caption{Performance of \tool{} evaluated on watermarked images (\dwm) generated at Stage 2 under different distortion types.}
\setlength{\tabcolsep}{3pt}
\footnotesize
\begin{tabular}{lccccc}
\toprule
\textbf{Distortion} & \textbf{\wmacc (\%)} & \textbf{\bitacc (\%)} & \textbf{P (\%)} & \textbf{R (\%)} & \textbf{F1 (\%)} \\
\midrule
Identity        & 99.8 & 97.2 & 99.6 & 100 & 99.8 \\
Blur            & 97.6 & 98.0 & 95.4 & 100 & 97.7 \\
Noise           & 99.8 & 98.5 & 99.6 & 100 & 99.8 \\
JPEG            & 99.8 & 97.2 & 99.6 & 100 & 99.8 \\
Brightness      & 99.8 & 96.3 & 99.6 & 100 & 99.8 \\
Contrast        & 99.8 & 96.5 & 99.6 & 100 & 99.8 \\
Resize          & 96.6 & 98.1 & 93.6 & 100 & 96.7 \\
\midrule
\textbf{Average} & \textbf{99.0} & \textbf{97.4} & \textbf{98.1} & \textbf{100.0} & \textbf{99.1} \\
\bottomrule
\end{tabular}
\label{tab:distortion}
\end{table}

\begin{table}[th!]
\centering
\caption{Performance of \tool{} evaluated on downstream DDPM's generated outputs (\dadv) at Stage 4 under different distortions. The DDPM is trained with the default UNet finetuning.}
\setlength{\tabcolsep}{3pt}
\footnotesize
\begin{tabular}{lccccc}
\toprule
\textbf{Distortion} & \textbf{\wmacc (\%)} & \textbf{\bitacc (\%)} & \textbf{P (\%)} & \textbf{R (\%)} & \textbf{F1 (\%)} \\
\midrule
Identity        & 99.5 & 84.4 & 99.0 & 100 & 99.5 \\
Gaussian Noise  & 98.5 & 85.5 & 97.1 & 100 & 98.5 \\
Brightness      & 99.0 & 84.3 & 98.0 & 100 & 99.0 \\
Contrast        & 99.5 & 85.1 & 99.0 & 100 & 99.5 \\
Blur            & 98.5 & 86.3 & 97.1 & 100 & 98.5 \\
JPEG            & 99.0 & 85.4 & 98.0 & 100 & 99.0 \\
Resize          & 98.5 & 87.3 & 97.1 & 100 & 98.5 \\
\midrule
\textbf{Average} & \textbf{98.93} & \textbf{85.2} & \textbf{97.9} & \textbf{100} & \textbf{98.9} \\
\bottomrule
\end{tabular}
\label{tab:radio_distortion}
\end{table}

\section{Common Distortions in Evaluation}\label{app:distortion}
As discussed in Section~\ref{subsec:robust}, we evaluate the resistance of our watermarking scheme against a wide range of distortions. 
For blur perturbations, we use Gaussian kernels with standard deviations ranging from 0.3 to 0.8. Additive Gaussian noise is introduced with standard deviations ranging between 0.005 and 0.02, while JPEG-like compression artifacts are simulated using random multiplicative noise factors in the range of 0.01–0.12. Spatial distortions are tested by resizing images with scale factors from 0.85 to 0.95, followed by bilinear interpolation back to the original resolution. Photometric perturbations such as brightness and contrast are adjusted with random multiplicative factors sampled from 0.9–1.1. These perturbations simulate the types of degradations that commonly occur in real-world imaging pipelines, enabling a systematic evaluation of the model’s resistance to visual noise and distortions.

Table~\ref{tab:distortion} presents \tool's watermark embedding performance and resistance to distortions at Stage 2, directly after watermark embedding. Table~\ref{tab:radio_distortion} records the performance of \tool when the finetuning script is the default UNet, complementary to Table~\ref{tab:distortion_8_lora_32} where the finetuning script is LoRA with $\texttt{rank}=32$.

\begin{table}[th!]
\centering
\caption{Performance of Recipe evaluated on watermarked images at Stage 2 under different distortion types.}
\setlength{\tabcolsep}{3pt}
\footnotesize
\begin{tabular}{lccccc}
\toprule
\textbf{Distortion} & \textbf{\wmacc (\%)} & \textbf{\bitacc (\%)} & \textbf{P (\%)} & \textbf{R (\%)} & \textbf{F1 (\%)} \\
\midrule
Identity        & 99.4 & 57.0 & 98.8 & 100 & 99.4 \\
Gaussian Noise  & 99.2 & 57.0 & 98.5 & 100 & 99.3 \\
Brightness      & 99.0 & 57.5 & 98.0 & 100 & 99.0 \\
Contrast        & 99.2 & 57.3 & 98.4 & 100 & 99.2 \\
Blur            & 99.4 & 57.0 & 98.8 & 100 & 99.4 \\
JPEG            & 99.2 & 57.6 & 98.3 & 100 & 99.2 \\
Resize          & 99.4 & 57.0 & 98.8 & 100 & 99.4 \\
\midrule
\textbf{Average} & \textbf{99.26} & \textbf{57.2} & \textbf{98.5} & \textbf{100} & \textbf{99.27} \\
\bottomrule
\end{tabular}
\label{tab:recipe_stage2}
\end{table}

\begin{table}[th!]
\centering
\caption{Performance of Recipe evaluated on downstream DDPM's generated outputs at Stage 4 under different distortion types. The DDPM is trained with the default UNet finetuning script.}
\setlength{\tabcolsep}{3pt}
\footnotesize
\begin{tabular}{lccccc}
\toprule
\textbf{Distortion} & \textbf{\wmacc (\%)} & \textbf{\bitacc (\%)} & \textbf{P (\%)} & \textbf{R (\%)} & \textbf{F1 (\%)} \\
\midrule
Identity        & 50.5 & 15.9 & 66.7 & 2.0 & 3.9 \\
Gaussian Noise  & 51.0 & 16.0 & 100.0 & 2.0 & 3.9 \\
Brightness      & 51.5 & 18.4 & 80.0 & 4.0 & 7.6 \\
Contrast        & 51.0 & 17.9 & 66.7 & 4.0 & 7.5 \\
Blur            & 50.5 & 15.8 & 66.7 & 2.0 & 3.9 \\
JPEG            & 50.5 & 17.2 & 60.0 & 3.0 & 5.7 \\
Resize          & 50.5 & 15.9 & 66.7 & 2.0 & 3.9 \\
\midrule
\textbf{Average} & \textbf{50.79} & \textbf{16.73} & \textbf{72.4} & \textbf{2.71} & \textbf{5.2} \\
\bottomrule
\end{tabular}
\label{tab:recipe_stage4}
\end{table}

\begin{table}[th!]
\centering
\caption{Performance of Recipe evaluated on downstream DDPM's generated outputs at Stage 4 under different distortion types. The DDPM is trained with LoRA finetuning script.}
\setlength{\tabcolsep}{3pt}
\footnotesize
\begin{tabular}{lccccc}
\toprule
\textbf{Distortion} & \textbf{\wmacc (\%)} & \textbf{\bitacc (\%)} & \textbf{P (\%)} & \textbf{R (\%)} & \textbf{F1 (\%)} \\
\midrule
Identity        & 50.5 & 15.9 & 66.7 & 2.0 & 3.9 \\
Gaussian Noise  & 49.5 & 16.2 & 40.0 & 2.0 & 3.8 \\
Brightness      & 51.0 & 15.3 & 100.0 & 2.0 & 3.9 \\
Contrast        & 51.0 & 15.1 & 100.0 & 2.0 & 3.9 \\
Blur            & 51.0 & 15.7 & 100.0 & 2.0 & 3.9 \\
JPEG            & 50.5 & 18.5 & 66.7 & 2.0 & 3.9 \\
Resize          & 50.5 & 15.9 & 66.7 & 2.0 & 3.9 \\
\midrule
\textbf{Average} & \textbf{50.57} & \textbf{16.09} & \textbf{77.16} & \textbf{2.00} & \textbf{3.89} \\
\bottomrule
\end{tabular}
\label{tab:recipe_stage4_lora}
\end{table}

\section{Detailed Evaluation Results of Recipe}\label{app:recipe}
Complementary to Section~\ref{subsec:radioactive}, we present the baseline competitor Recipe's~\cite{zhao2023recipe} performance at Stage 2 in Table~\ref{tab:recipe_stage2} and its performance at Stage 4 in Table~\ref{tab:recipe_stage4} for default UNet finetuning and Table~\ref{tab:recipe_stage4_lora} for LoRA finetuning.

\begin{table}[H]
\centering
\footnotesize
\caption{
Performance of \tool{} when downstream DDPM finetunes using LoRA with different ranks at step 9{,}000 and 8-bit secret size.
}
\begin{tabular}{lccccc}
\toprule
\textbf{Rank} & \textbf{\wmacc (\%)} & \textbf{\bitacc (\%)} & \textbf{P (\%)} & \textbf{R (\%)} & \textbf{F1 (\%)} \\
\midrule
8 & 98.29 & 89.29 & 97.34 & 99.29 & 98.30 \\
16 & 98.14  & 91.14 & 97.20 & 99.14 & 98.16 \\
32 & 98.57 & 95.07 & 97.22 & 100.00 & 98.59 \\
\bottomrule
\end{tabular}
\label{tab:lora_rank_compare}
\end{table}

\begin{table}[ht!]
\centering
\setlength{\tabcolsep}{3pt}
\footnotesize
\caption{
Performance of \tool{} evaluated on downstream DDPM's generated outputs (\dadv) at Stage 4 under different distortion types. The DDPM is trained with the LoRA finetuning script ($\texttt{rank} = 16$). Secret bit size is 8.
}
\begin{tabular}{lccccc}
\toprule
\textbf{Distortion} & \textbf{\wmacc (\%)} & \textbf{\bitacc (\%)} & \textbf{P (\%)} & \textbf{R (\%)} & \textbf{F1 (\%)} \\
\midrule
Identity        & 98.50 & 89.12 & 98.02 & 99.00 & 98.51 \\
Gaussian Noise  & 98.00 & 92.88 & 97.06 & 99.00 & 98.02 \\
Brightness      & 98.50 & 90.00 & 98.02 & 99.00 & 98.51 \\
Contrast        & 98.50 & 88.75 & 98.02 & 99.00 & 98.51 \\
Blur            & 98.50 & 92.50 & 98.02 & 99.00 & 98.51 \\
JPEG            & 99.00 & 91.12 & 98.04 & 100.00 & 99.01 \\
Resize          & 96.00 & 94.62 & 93.40 & 99.00 & 96.12 \\
\midrule
\textbf{Average} & \textbf{98.14} & \textbf{91.14} & \textbf{97.20} & \textbf{99.14} & \textbf{98.16} \\
\bottomrule
\end{tabular}
\label{tab:distortion_8_lora_16}
\end{table}

\begin{table}[ht!]
\centering
\setlength{\tabcolsep}{3pt}
\footnotesize
\caption{
Performance of \tool{} evaluated on downstream DDPM's generated outputs at Stage 4 under different distortion types. The DDPM is trained with the LoRA finetuning script ($\texttt{rank} = 8$). Secret bit size is 8.
}
\begin{tabular}{lccccc}
\toprule
\textbf{Distortion} & \textbf{\wmacc (\%)} & \textbf{\bitacc (\%)} & \textbf{P (\%)} & \textbf{R (\%)} & \textbf{F1 (\%)} \\
\midrule
Identity        & 98.5 & 87.62 & 98.02 & 99.00 & 98.51 \\
Gaussian Noise  & 99.0 & 88.75 & 98.04 & 100.00 & 99.01 \\
Brightness      & 98.0 & 87.38 & 97.06 & 99.00 & 98.02 \\
Contrast        & 98.5 & 87.00 & 98.02 & 99.00 & 98.51 \\
Blur            & 98.5 & 92.12 & 98.02 & 99.00 & 98.51 \\
JPEG            & 98.5 & 88.62 & 98.02 & 99.00 & 98.51 \\
Resize          & 97.0 & 94.75 & 94.34 & 100.00 & 97.09 \\
\midrule
\textbf{Average} & \textbf{98.29} & \textbf{89.29} & \textbf{97.34} & \textbf{99.29} & \textbf{98.30} \\
\bottomrule
\end{tabular}
\label{tab:distortion_8_lora_8}
\end{table}

\section{LoRA Rank Analysis}\label{app:lora_rank}
Table~\ref{tab:lora_rank_compare} shows the comparison of \tool's performance at Stage 4 when finetuned for 9{,}000 steps using LoRA with different ranks (32, 16, 8). Detailed results on resistance to distortions are presented in Tables~\ref{tab:distortion_8_lora_32}, \ref{tab:distortion_8_lora_16}, and \ref{tab:distortion_8_lora_8}. 
While \wmacc for all tested LoRA ranks are similar (\ie, above 98\%), a higher LoRA rank leads to a higher \bitacc. However, it is worth noting that \bitacc for all tested ranks reach 94.50\% eventually with further finetuning (See Table~\ref{tab:finetune_steps_lora_all}).

\begin{figure*}[ht!]
\centering

\begin{minipage}[b]{0.32\textwidth}
    \centering
    \subcaption*{\textbf{Original} \hspace{1.5cm} \textbf{Watermarked}}
    \includegraphics[width=0.49\textwidth]{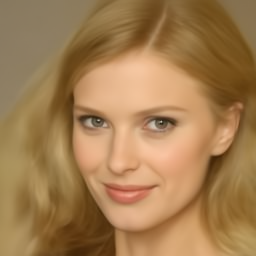}
    \includegraphics[width=0.49\textwidth]{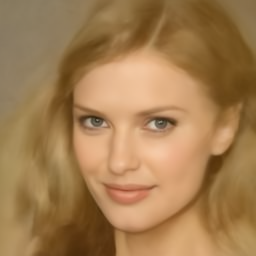}
\end{minipage}
\hfill
\begin{minipage}[b]{0.32\textwidth}
    \centering
    \subcaption*{\textbf{Original} \hspace{1.5cm} \textbf{Watermarked}}
    \includegraphics[width=0.49\textwidth]{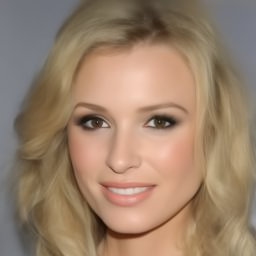}
    \includegraphics[width=0.49\textwidth]{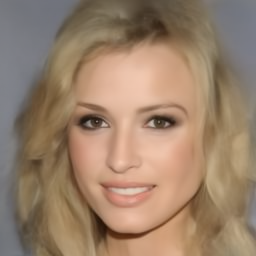}
\end{minipage}
\hfill
\begin{minipage}[b]{0.32\textwidth}
    \centering
    \subcaption*{\textbf{Original} \hspace{1.5cm} \textbf{Watermarked}}
    \includegraphics[width=0.49\textwidth]{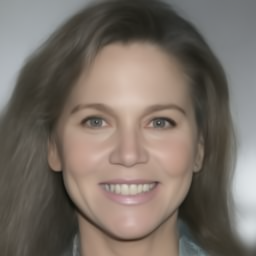}
    \includegraphics[width=0.49\textwidth]{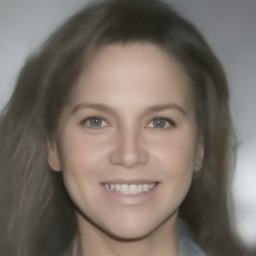}
\end{minipage}

\begin{minipage}[b]{0.32\textwidth}
    \centering
    \includegraphics[width=0.49\textwidth]{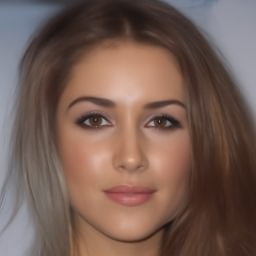}
    \includegraphics[width=0.49\textwidth]{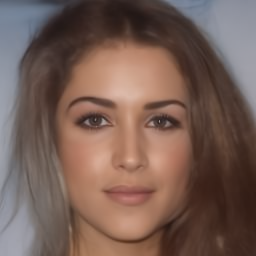}
\end{minipage}
\hfill
\begin{minipage}[b]{0.32\textwidth}
    \centering
    \includegraphics[width=0.49\textwidth]{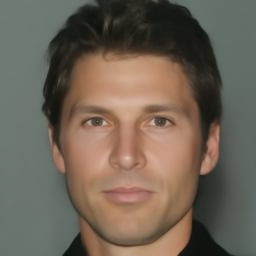}
    \includegraphics[width=0.49\textwidth]{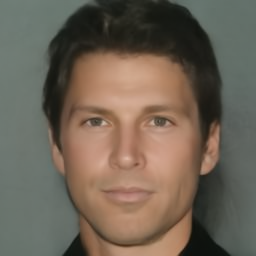}
\end{minipage}
\hfill
\begin{minipage}[b]{0.32\textwidth}
    \centering
    \includegraphics[width=0.49\textwidth]{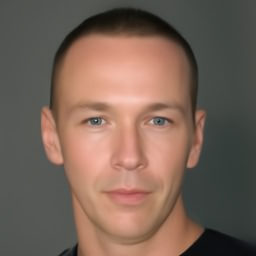}
    \includegraphics[width=0.49\textwidth]{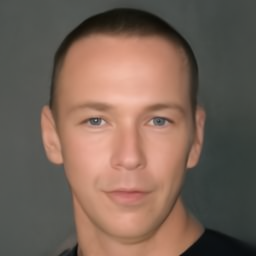}
\end{minipage}

\begin{minipage}[b]{0.32\textwidth}
    \centering
    \includegraphics[width=0.49\textwidth]{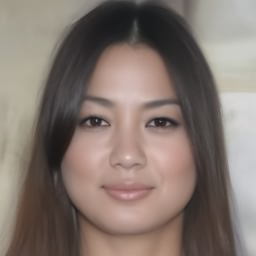}
    \includegraphics[width=0.49\textwidth]{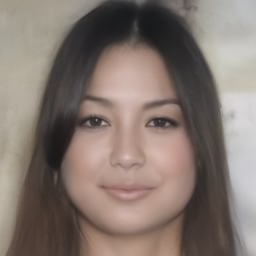}
\end{minipage}
\hfill
\begin{minipage}[b]{0.32\textwidth}
    \centering
    \includegraphics[width=0.49\textwidth]{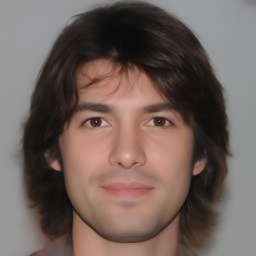}
    \includegraphics[width=0.49\textwidth]{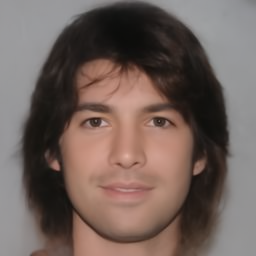}
\end{minipage}
\hfill
\begin{minipage}[b]{0.32\textwidth}
    \centering
    \includegraphics[width=0.49\textwidth]{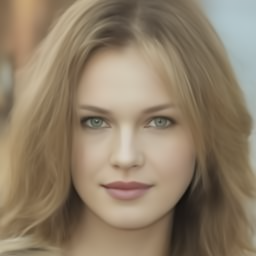}
    \includegraphics[width=0.49\textwidth]{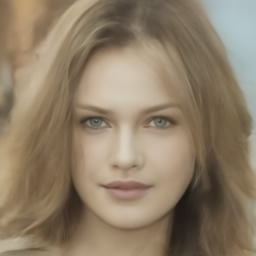}
\end{minipage}

\begin{minipage}[b]{0.32\textwidth}
    \centering
    \includegraphics[width=0.49\textwidth]{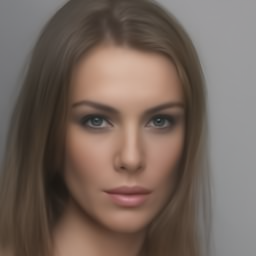}
    \includegraphics[width=0.49\textwidth]{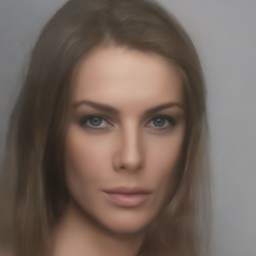}
\end{minipage}
\hfill
\begin{minipage}[b]{0.32\textwidth}
    \centering
    \includegraphics[width=0.49\textwidth]{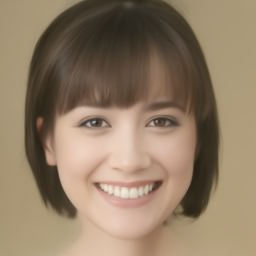}
    \includegraphics[width=0.49\textwidth]{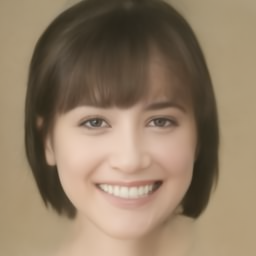}
\end{minipage}
\hfill
\begin{minipage}[b]{0.32\textwidth}
    \centering
    \includegraphics[width=0.49\textwidth]{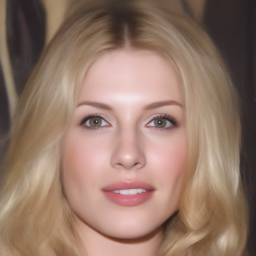}
    \includegraphics[width=0.49\textwidth]{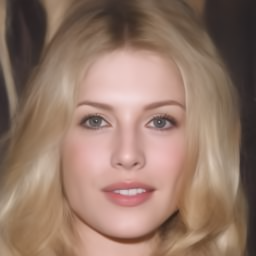}
\end{minipage}

\caption{Visual comparison between original and watermarked images across multiple samples. Each pair shows the original (left) and its corresponding watermarked (right) image.}
\label{fig:appendix_orig_wm_grid}
\end{figure*}

\section{Sample Pairs of Original and Watermarked Images} \label{app:invisible}
We present additional pairs of original and watermarked images generated by \tool in Figure~\ref{fig:appendix_orig_wm_grid}.

\begin{figure*}[ht!]
\centering
\setlength{\tabcolsep}{2.5pt} %
\renewcommand{\arraystretch}{0}
\begin{tabular}{c*{6}{c}}
 & \textbf{500} & \textbf{1k} & \textbf{3k} & \textbf{6k} & \textbf{9k} & \textbf{10k} \\
[4pt] %
\centering\rotatebox{90}{\small \dorig} &
\includegraphics[width=0.15\textwidth]{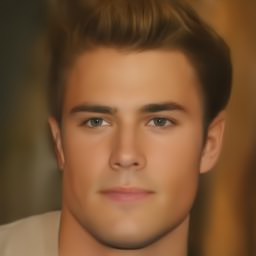} &
\includegraphics[width=0.15\textwidth]{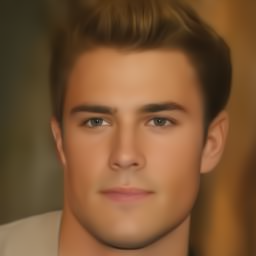} &
\includegraphics[width=0.15\textwidth]{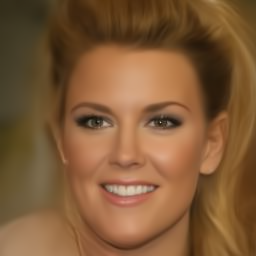} &
\includegraphics[width=0.15\textwidth]{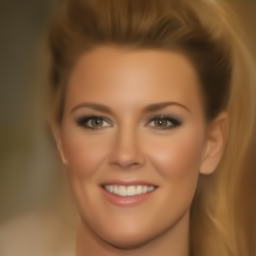} &
\includegraphics[width=0.15\textwidth]{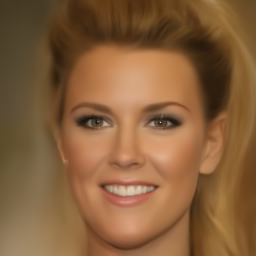} &
\includegraphics[width=0.15\textwidth]{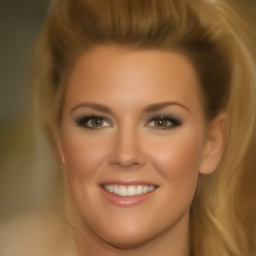} \\
[4pt]
\centering\rotatebox{90}{\small \dwm~(8\text{-}bit)} &
\includegraphics[width=0.15\textwidth]{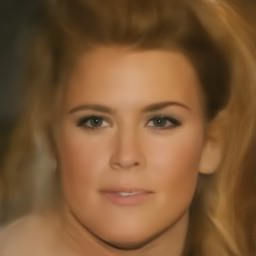} &
\includegraphics[width=0.15\textwidth]{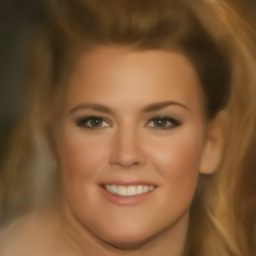} &
\includegraphics[width=0.15\textwidth]{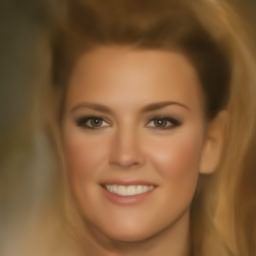} &
\includegraphics[width=0.15\textwidth]{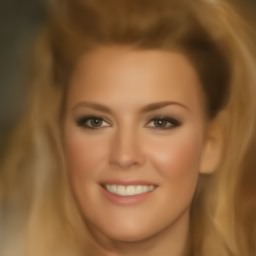} &
\includegraphics[width=0.15\textwidth]{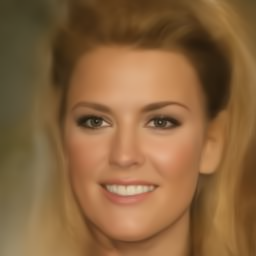} &
\includegraphics[width=0.15\textwidth]{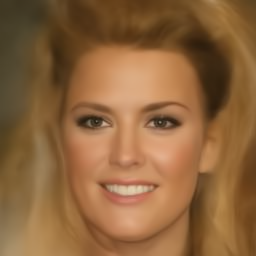} \\
[4pt]
\centering\rotatebox{90}{\small \dwm~(16\text{-}bit)} &
\includegraphics[width=0.15\textwidth]{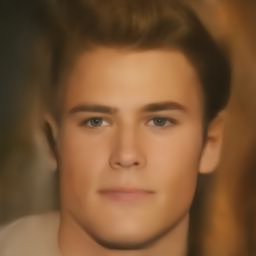} &
\includegraphics[width=0.15\textwidth]{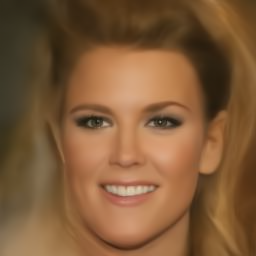} &
\includegraphics[width=0.15\textwidth]{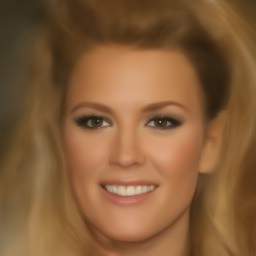} &
\includegraphics[width=0.15\textwidth]{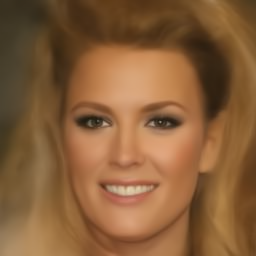} &
\includegraphics[width=0.15\textwidth]{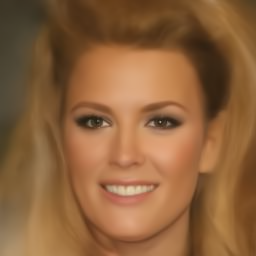} &
\includegraphics[width=0.15\textwidth]{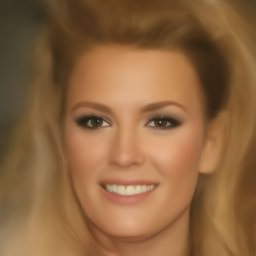} \\
[4pt]

\centering\rotatebox{90}{\small \dwm~(32\text{-}bit)} &
\includegraphics[width=0.15\textwidth]{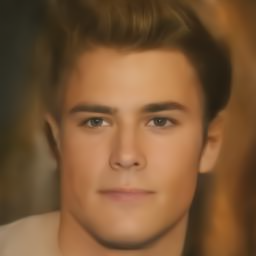} &
\includegraphics[width=0.15\textwidth]{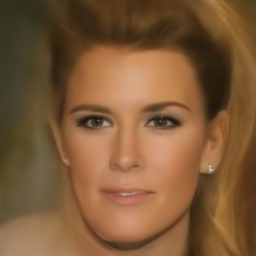} &
\includegraphics[width=0.15\textwidth]{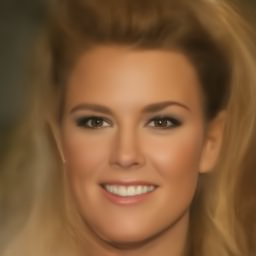} &
\includegraphics[width=0.15\textwidth]{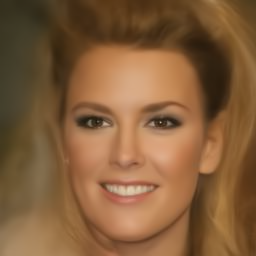} &
\includegraphics[width=0.15\textwidth]{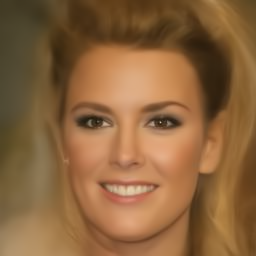} &
\includegraphics[width=0.15\textwidth]{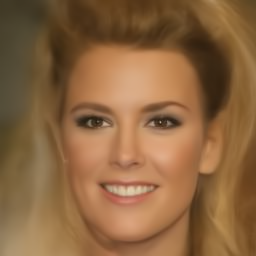} \\
\end{tabular}
\caption{
Comparison of generations from \madv finetuned with LoRA ($\texttt{rank}=32$) across finetuning steps when trained on (top) \dorig, (second) \dwm with 8-bit secret, (third) \dwm with 16-bit secret, and (bottom) \dwm with 32-bit secret.}
\label{fig:generated_steps_lora32_compare}
\end{figure*}

\section{Visual Results as Finetuning on Clean Images versus Finetuning on Watermarked Images}\label{app:semantic_shift}
Figure~\ref{fig:generated_steps_lora32_compare} shows the comparison of semantic shifts and visual quality of \dclean and \dadv generated by \madv LoRA-finetuned on 10k clean (\dorig), 8-bit, 16-bit, and 32-bit secret embedded watermarked data samples (\dwm).

\begin{table}[ht!]
\centering
\footnotesize
\setlength{\tabcolsep}{2pt}
\caption{
Impact of the size of \dwm acquired by the adversary to finetune \madv on \tool's performance. \madv is LoRA ($\texttt{rank}=32$) finetuned on \dwm with an 8-bit secret embedded.
}
\begin{tabular}{llcccccc}
\toprule
\textbf{$|$\dwm$|$} & \textbf{\wmacc (\%)} & \textbf{\bitacc (\%)} & \textbf{P (\%)} & \textbf{R (\%)} & \textbf{F1 (\%)} \\
\midrule
1{,}000   & 98.57 & 97.29 & 97.22 & 100 & 98.59 \\
3{,}000   & 98.64 & 97.29 & 97.36 & 100 & 98.66 \\
5{,}000   & 98.50 & 94.86 & 97.22 & 99.86 & 98.52 \\
7{,}000   & 98.43 & 94.21 & 97.08 & 99.86 & 98.45 \\
10{,}000  & 98.57 & 95.07 & 97.22 & 100 & 98.59 \\
15{,}000  & 98.57 & 95.43 & 97.22 & 100 & 98.59 \\
20{,}000  & 98.57 & 96.43 & 97.22 & 100 & 98.59 \\
\bottomrule
\end{tabular}
\label{tab:num_training_steps}
\end{table}

\begin{table}[ht!]
\centering
\footnotesize
\caption{
\tool{}'s radioactivity performance on downstream DDPM \madv finetuned with the default UNet script using different secret bit lengths.
}
\begin{tabular}{lccccc}
\toprule
\textbf{\# Bits} & \textbf{\wmacc (\%)} & \textbf{\bitacc (\%)} & \textbf{P (\%)} & \textbf{R (\%)} & \textbf{F1 (\%)} \\
\midrule
8   & 98.14 & 87.43 & 97.33 & 99.00 & 98.16 \\
16  & 99.21 & 78.93 & 98.45 & 100   & 99.22 \\
32  & 93.50 & 68.50 & 88.50 & 100 & 93.90 \\
\bottomrule
\end{tabular}
\label{tab:bits}
\end{table}

\begin{table*}[th!]
\centering
\setlength{\tabcolsep}{6pt}
\footnotesize
\caption{
Performance of \tool{} evaluated on downstream DDPM (\madv)'s generated outputs (\dadv) at Stage 4 under different finetuning scripts (\texttt{UNet} vs. \texttt{LoRA}) and bit configurations. The results are collected at the finetuning step 9{,}000.
}
\begin{tabular}{lllccccc}
\toprule
\textbf{Finetuning Script} & \textbf{\# Bits} & \textbf{Distortion} & \textbf{\wmacc (\%)} & \textbf{\bitacc (\%)} & \textbf{P (\%)} & \textbf{R (\%)} & \textbf{F1 (\%)} \\
\midrule
\multirow{25}{*}{\textbf{UNet}} 
 & \multirow{8}{*}{8}
 & Identity        & 99.5 & 84.4 & 99.0 & 100 & 99.5 \\
& &Gaussian Noise  & 98.5 & 85.5 & 97.1 & 100 & 98.5 \\
& &Brightness      & 99.0 & 84.3 & 98.0 & 100 & 99.0 \\
& &Contrast        & 99.5 & 85.1 & 99.0 & 100 & 99.5 \\
& &Blur            & 98.5 & 86.3 & 97.1 & 100 & 98.5 \\
&& JPEG            & 99.0 & 85.4 & 98.0 & 100 & 99.0 \\
& &Resize          & 98.5 & 87.3 & 97.1 & 100 & 98.5 \\
& & \textbf{Average} & \textbf{98.93} & \textbf{85.2} & \textbf{97.9} & \textbf{100} & \textbf{98.9} \\
\cmidrule(lr){2-8}
 & \multirow{8}{*}{16} 
 & Identity        & 99.5 & 78.50 & 99.01 & 100.00 & 99.50 \\
 &  & Gaussian Noise  & 99.5 & 80.12 & 99.01 & 100.00 & 99.50 \\
 &  & Brightness      & 99.5 & 77.25 & 99.01 & 100.00 & 99.50 \\
 &  & Contrast        & 99.5 & 80.12 & 99.01 & 100.00 & 99.50 \\
 &  & Blur            & 99.0 & 78.75 & 98.04 & 100.00 & 99.01 \\
 &  & JPEG            & 99.5 & 79.94 & 99.01 & 100.00 & 99.50 \\
 &  & Resize          & 98.0 & 79.12 & 96.15 & 100.00 & 98.04 \\
 &  & \textbf{Average} & \textbf{99.21} & \textbf{78.93} & \textbf{98.45} & \textbf{100.00} & \textbf{99.22} \\
\cmidrule(lr){2-8}
 & \multirow{8}{*}{32} 
 & Identity        & 93.5 & 68.38 & 91.43 & 96.00 & 93.66 \\
 &  & Gaussian Noise  & 91.5 & 68.34 & 88.79 & 95.00 & 91.79 \\
 &  & Brightness      & 93.5 & 68.22 & 93.07 & 94.00 & 93.53 \\
 &  & Contrast        & 92.0 & 68.44 & 89.62 & 95.00 & 92.23 \\
 &  & Blur            & 89.5 & 68.41 & 84.96 & 96.00 & 90.14 \\
 &  & JPEG            & 92.0 & 68.34 & 89.62 & 95.00 & 92.23 \\
 &  & Resize          & 86.5 & 68.44 & 80.17 & 97.00 & 87.78 \\
 &  & \textbf{Average} & \textbf{91.21} & \textbf{68.00} & \textbf{88.01} & \textbf{95.43} & \textbf{91.57} \\
\midrule
\multirow{25}{*}{\textbf{LoRA} ($\texttt{rank}=32$)} 
 & \multirow{8}{*}{8}
& Identity        & 99.00 & 94.00 & 98.04 & 100.00 & 99.01 \\
&& Gaussian Noise  & 99.00 & 95.75 & 98.04 & 100.00 & 99.01 \\
&& Brightness      & 98.50 & 94.38 & 97.09 & 100.00 & 98.52 \\
& &Contrast        & 99.00 & 94.50 & 98.04 & 100.00 & 99.01 \\
& &Blur            & 98.50 & 95.50 & 97.09 & 100.00 & 98.52 \\
& &JPEG            & 99.00 & 95.25 & 98.04 & 100.00 & 99.01 \\
&& Resize          & 97.00 & 97.12 & 94.34 & 100.00 & 97.09 \\
& & \textbf{Average} & \textbf{98.57} & \textbf{95.07} & \textbf{97.22} & \textbf{100.00} & \textbf{98.59} \\
\cmidrule(lr){2-8}
 & \multirow{8}{*}{16} 
 & Identity        & 99.5 & 92.88 & 100.00 & 99.00 & 99.50 \\
 &  & Gaussian Noise  & 99.5 & 94.12 & 100.00 & 99.00 & 99.50 \\
 &  & Brightness      & 100.0 & 93.94 & 100.00 & 100.00 & 100.00 \\
 &  & Contrast        & 99.5 & 93.50 & 100.00 & 99.00 & 99.50 \\
 &  & Blur            & 99.5 & 93.81 & 100.00 & 99.00 & 99.50 \\
 &  & JPEG            & 100.0 & 93.81 & 100.00 & 100.00 & 100.00 \\
 &  & Resize          & 98.5 & 94.00 & 98.02 & 99.00 & 98.51 \\
 &  & \textbf{Average} & \textbf{99.50} & \textbf{93.73} & \textbf{99.43} & \textbf{99.14} & \textbf{99.50} \\
\cmidrule(lr){2-8}
 & \multirow{8}{*}{32} 
 & Identity        & 91.0 & 68.81 & 89.42 & 93.00 & 91.18 \\
 &  & Gaussian Noise  & 92.0 & 68.72 & 89.62 & 95.00 & 92.23 \\
 &  & Brightness      & 91.0 & 68.59 & 89.42 & 93.00 & 91.18 \\
 &  & Contrast        & 89.0 & 68.53 & 87.50 & 91.00 & 89.22 \\
 &  & Blur            & 90.5 & 68.78 & 85.84 & 97.00 & 91.08 \\
 &  & JPEG            & 92.5 & 68.69 & 89.72 & 96.00 & 92.75 \\
 &  & Resize          & 89.0 & 68.78 & 83.62 & 97.00 & 89.81 \\
 &  & \textbf{Average} & \textbf{90.71} & \textbf{68.70} & \textbf{87.88} & \textbf{94.57} & \textbf{91.04} \\
\bottomrule
\end{tabular}
\label{tab:combined_distortion}
\end{table*}

\begin{table}[th!]
\centering
\footnotesize
\caption{
Performance of \tool{} at different finetuning steps when \madv is finetuned with default UNet finetuning script.
}
\begin{tabular}{lccccc}
\toprule
\textbf{Steps} & \textbf{\wmacc (\%)} & \textbf{\bitacc (\%)} & \textbf{P (\%)} & \textbf{R (\%)} & \textbf{F1 (\%)} \\
\midrule
1{,}000  & 97.79 & 62.86 & 97.85 & 97.71 & 97.78 \\
3{,}000  & 97.00 & 61.14 & 97.82 & 96.14 & 96.97 \\
5{,}000  & 98.50 & 81.86 & 97.88 & 99.14 & 98.51 \\
6{,}000  & 98.93 & 85.21 & 97.90 & 100.00 & 98.94 \\
9{,}000  & 98.14 & 87.43 & 97.33 & 99.00 & 98.16 \\
10{,}000 & 98.57 & 86.43 & 98.02 & 99.14 & 98.58 \\
20{,}000 & 98.50 & 86.79 & 98.02 & 99.00 & 98.51 \\
30{,}000 & 99.00 & 87.64 & 98.04 & 100.00 & 99.01 \\
\bottomrule
\end{tabular}
\label{tab:finetune_steps}
\end{table}

\begin{table}
\centering
\footnotesize
\setlength{\tabcolsep}{3pt}
\caption{
Performance of \tool{} at different finetuning steps when \madv is LoRA-finetuned ($\texttt{ranks} =  32, 16, 8$).
}
\begin{tabular}{llccccc}
\toprule
\textbf{Rank} & \textbf{Steps} & \textbf{\wmacc (\%)} & \textbf{\bitacc (\%)} & \textbf{P (\%)} & \textbf{R (\%)} & \textbf{F1 (\%)} \\
\midrule
\multirow{8}{*}{32}
& 500     & 94.21 & 50.07 & 96.82 & 91.43 & 94.05 \\
& 1{,}000 & 96.64 & 65.71 & 97.11 & 96.14 & 96.63 \\
& 2{,}000 & 97.57 & 83.71 & 97.03 & 98.14 & 97.59 \\
& 3{,}000 & 97.93 & 86.36 & 97.19 & 98.71 & 97.94 \\
& 5{,}000 & 98.43 & 90.71 & 97.08 & 99.86 & 98.45 \\
& 6{,}000 & 98.50 & 90.86 & 97.09 & 100.00 & 98.52 \\
& 9{,}000 & 98.57 & 95.07 & 97.22 & 100.00 & 98.59 \\
& 10{,}000& 98.57 & 94.50 & 97.22 & 100.00 & 98.59 \\
\midrule
\multirow{9}{*}{16}
& 500     & 83.29 & 22.07 & 96.23 & 69.29 & 80.56 \\
& 1{,}000 & 93.43 & 44.64 & 96.91 & 89.71 & 93.18 \\
& 2{,}000 & 97.00 & 70.79 & 97.27 & 96.71 & 96.99 \\
& 3{,}000 & 97.64 & 78.29 & 97.30 & 98.00 & 97.65 \\
& 5{,}000 & 98.00 & 84.93 & 97.32 & 98.71 & 98.01 \\
& 6{,}000 & 97.79 & 87.50 & 97.31 & 98.29 & 97.80 \\
& 9{,}000 & 98.14 & 91.14 & 97.20 & 99.14 & 98.16 \\
& 10{,}000& 98.64 & 91.93 & 97.36 & 100.00 & 98.66 \\
& 30{,}000& 98.43 & 95.64 & 97.35 & 99.57 & 98.45 \\
\midrule
\multirow{9}{*}{8}
& 500     & 90.36 & 35.21 & 96.85 & 83.43 & 89.64 \\
& 1{,}000 & 95.57 & 57.14 & 97.19 & 93.86 & 95.49 \\
& 2{,}000 & 98.07 & 79.36 & 97.19 & 99.00 & 98.09 \\
& 3{,}000 & 97.79 & 81.93 & 97.05 & 98.57 & 97.80 \\
& 5{,}000 & 98.43 & 85.57 & 97.21 & 99.71 & 98.45 \\
& 6{,}000 & 98.43 & 89.50 & 97.35 & 99.57 & 98.45 \\
& 9{,}000 & 98.29 & 89.29 & 97.34 & 99.29 & 98.30 \\
& 10{,}000 & 98.57 & 94.50 & 97.35 & 99.86 & 98.59 \\
\bottomrule
\end{tabular}
\label{tab:finetune_steps_lora_all}
\end{table}

\section{Tabularized Results on Relationship between \tool's Effectiveness and Number of Training Samples}\label{app:training_steps}
Table~\ref{tab:num_training_steps} shows the detailed performance of \tool with different \dwm sizes, which is complementary to the discussion in Section~\ref{subsec:radioactive}.

\section{Tabularized Results on Secret Size Analysis}\label{app:sec_size}
As Table~\ref{tab:bits_lora} in Section~\ref{subsec:sec_size} presents, LoRA has higher \bitacc as we increase the secret size. It achieves above 93\% \bitacc when the secret size is less than 32 bits and drops to 68\% when 32 bits are embedded.
Table~\ref{tab:bits} shows the average radioactivity performance of \tool when \madv is finetuned with the default UNet script.
Table~\ref{tab:combined_distortion} illustrates the effectiveness of \tool in detail when embedding 8-bit, 16-bit, and 32-bit secrets and finetuning using the default UNet script and LoRA. The results match the intuition discussed in Section~\ref{subsec:theory} that as more bits are transmitted during downstream finetuning, \bitacc drops gracefully. This motivates the integration of error correction compatibility in our tool (See Appendix~\ref{app:bch}).

\section{Tabularized Results on Relationship between Secret Recovery Accuracy and Finetuning} \label{app:discuss_steps}
Tables \ref{tab:finetune_steps} and~\ref{tab:finetune_steps_lora_all} show how \tool’s performance evolves across downstream DDPM finetuning steps using the default finetuning script and LoRA.
When \madv is finetuned with the default UNet finetuning script, the trend matches the discussion about LoRA finetuning in Section~\ref{subsec:acc_steps}.
At the early stages (1k–3k steps), the \bitacc is relatively low (61–63\%), indicating that the watermark signal has not yet fully stabilized.
As default UNet finetuning progresses, both overall accuracy and secret accuracy steadily improve, reaching over 85\% after 6k steps and exceeding 87\% by 30k steps, which matches our theoretical intuition in Section~\ref{subsec:theory}.
Precision, recall, and F1 remain consistently high throughout, suggesting that the watermark detector remains reliable.
These results demonstrate that a moderate number of finetuning iterations (6k-9k steps) is sufficient for convergence and high watermark recovery for both finetuning scripts.

\begin{table}[ht!]
\centering
\scriptsize
\caption{
Comparison of \tool{} performance on radioactivity to \madv with and without ECC for different watermark secret bit lengths. \madv finetuned with the default UNet finetuning script.
}
\setlength{\tabcolsep}{2pt}

\begin{tabular}{lllccccc}
\toprule
\textbf{ECC} & \textbf{\# Total} & \textbf{\# Sec.} & \textbf{\wmacc (\%)} & \textbf{\bitacc(\%)} & \textbf{P (\%)} & \textbf{R (\%)} & \textbf{F1 (\%)} \\
\midrule
None          & 8  & 8  & 98.14 & 87.43 & 97.33 & 99.00 & 98.16 \\
BCH(31, 8)    & 31 & 8  & 97.93 & 91.14 & 96.53 & 99.43 & 97.96 \\
\midrule
None          & 16 & 16 & 99.21 & 78.93 & 98.45 & 100.00 & 99.22 \\
BCH(31, 16)   & 31 & 16 & 99.43 & 93.02 & 99.01 & 99.86 & 99.43 \\
\bottomrule
\end{tabular}
\label{tab:ecc}
\end{table}

\begin{table}[th!]
\centering
\setlength{\tabcolsep}{3pt}
\footnotesize
\caption{
Performance of \tool{} evaluated on \madv's generated outputs at Stage 4 using \textbf{BCH(31, 8)} ECC under different distortions. \madv is finetuned with default UNet finetuning script.
}
\begin{tabular}{lccccc}
\toprule
\textbf{Distortion} & \textbf{\wmacc (\%)} & \textbf{\bitacc (\%)} & \textbf{P (\%)} & \textbf{R (\%)} & \textbf{F1 (\%)} \\
\midrule
Identity        & 98.5 & 89.88 & 98.02 & 99.00 & 98.51 \\
Gaussian Noise  & 98.0 & 92.75 & 97.06 & 99.00 & 98.02 \\
Brightness      & 99.5 & 88.88 & 99.01 & 100.00 & 99.50 \\
Contrast        & 99.0 & 89.75 & 99.00 & 99.00 & 99.00 \\
Blur            & 97.5 & 92.12 & 95.24 & 100.00 & 97.56 \\
JPEG            & 98.0 & 92.62 & 97.06 & 99.00 & 98.02 \\
Resize          & 95.0 & 93.50 & 90.91 & 100.00 & 95.24 \\
\midrule
\textbf{Average} & \textbf{97.93} & \textbf{91.14} & \textbf{96.53} & \textbf{99.43} & \textbf{97.96} \\
\bottomrule
\end{tabular}
\label{tab:bch_31_8}
\end{table}

\begin{table}[th!]
\centering
\setlength{\tabcolsep}{3pt}
\footnotesize
\caption{
Performance of \tool{} evaluated on \madv's generated outputs at Stage 4 using \textbf{BCH(31, 16)} ECC under different distortions. \madv is finetuned with the default UNet finetuning script.
}
\begin{tabular}{lccccc}
\toprule
\textbf{Distortion} & \textbf{\wmacc (\%)} & \textbf{\bitacc (\%)} & \textbf{P (\%)} & \textbf{R (\%)} & \textbf{F1 (\%)} \\
\midrule
Identity        & 99.5 & 90.88  & 99.01 & 100.00 & 99.50 \\
Blur            & 99.5 & 93.19 & 99.01 & 100.00 & 99.50 \\
Noise           & 99.0 & 92.31 & 99.00 & 99.00 & 99.00 \\
JPEG            & 99.5 & 93.06 & 99.01 & 100.00 & 99.50 \\
Brightness      & 100.0 & 94.06 & 100.00 & 100.00 & 100.00 \\
Contrast        & 99.5 & 93.12 & 99.01 & 100.00 & 99.50 \\
Resize          & 99.0 & 94.50 & 98.04 & 100.00 & 99.01 \\
\midrule
\textbf{Average} & \textbf{99.43} & \textbf{93.02} & \textbf{99.01} & \textbf{99.86} & \textbf{99.43} \\
\bottomrule
\end{tabular}
\label{tab:bch_31_16}
\end{table}

\section{BCH Error Correction Integration}\label{app:bch}
Motivated by the observations in the relationship between secret size and \bitacc when \madv is finetuned on the default UNet finetuning script with limited finetuning steps, \tool also supports Bose-Chaudhuri-Hocquenghem (BCH) error correction code (ECC)~\cite{BoseRayChaudhuri1960} to produce codewords $\mathbf{s}_{\text{ecc}} \in \{0,1\}^{n}$ with error correction capabilities. Specifically, the BCH($n$, $k$) encoding scheme allows the system to transmit parity bits to recover lost messages, where $n$ is the total codeword length and $k$ denotes the number of secret data bits. 
When ECC mode is enabled, the encoder $E_{\theta}$ operates on BCH-encoded secrets instead of raw $k$-bit secret vectors, while the detector adapts to predict $n$-bit codewords. During inference, the predicted codeword is BCH-decoded to reconstruct the original $k$-bit message.
We implement BCH using \texttt{bchlib}~\cite{bchlib}, which wraps the Linux kernel BCH library~\cite{linux_bch}. The implementation constrains the Galois field order to $m \in [5,15]$, corresponding to codeword lengths $n = 2^m - 1$ and setting the smallest supported $n$ to 31.
In practice, both BCH(31,8) and BCH(31,16) substantially improve robustness, increasing \bitacc{} beyond 90\% under downstream DM finetuning.
Apart from Table~\ref{tab:ecc}, which shows the comparison of \tool's performance with and without BCH error correction, Tables~\ref{tab:bch_31_8} and~\ref{tab:bch_31_16} show the detailed performance of \tool under various distortions when BCH is integrated. We draw the conclusion that BCH helps improve \bitacc to above 90\% even when more secret bits are embedded and transmitted.

\end{document}